\documentclass[12pt]{article}
\usepackage{axodraw}
\usepackage{amssymb}
\input xy
\xyoption{all}

\def\ll{_}
\def\uu{^}
\def\co{{\cal O}}
\newcommand{\heading}[1]{{\begin{center} {\it {#1} \rm} \end{center}}}
\def\b{\beta}
\def\d{\delta}
\def\bbb{\begin{eqnarray} }
\def\eee{\end{eqnarray}}
\def\cc{\,}
\def\hh{{1\over 2}}
\def\sqd{^2}
\def\llsk{\hspace{.2in}}
\def\L{\Lambda}
\def\pr{^\prime}
\def\nn{\nonumber}
\def\lrdd{\left (}
\def\rrdd{\right )}
\def\lsqq{\left [}
\def\rsqq{\right ]}
\def\s{\sigma}
\def\e{\epsilon}
\def\pp{\partial}
\def\dag{^\dagger}
\def\m{\mu}

\begin{document}

\hfill VT-IPNAS 10-20 

\hfill IPMU 10-0230

\vspace{1in}

\begin{center}

{\large\bf Sums over Topological Sectors }

{\large\bf and Quantization of Fayet-Iliopoulos Parameters}

\vspace{0.1in}

Simeon Hellerman$^1$ and Eric Sharpe$^2$

\begin{tabular}{cc}
{ \begin{tabular}{c}
$^1$ Institute for Physics and \\
\hspace*{0.3in} Mathematics of the Universe (IPMU) \\
5-1-5 Kashiwanoha \\
Kashiwa, 277-8583, Japan
\end{tabular} } 
&
{ \begin{tabular}{c}
$^2$ Physics Department\\
Virginia Tech \\
Blacksburg, VA  24061
\end{tabular} } 
\end{tabular}

{\tt simeon.hellerman.1@gmail.com}, 
{\tt ersharpe@vt.edu}\\

$\,$

\end{center}

In this paper we discuss quantization of the Fayet-Iliopoulos parameter
in supergravity theories with altered nonperturbative sectors,
which were recently used to argue a fractional
quantization condition.  Nonlinear sigma models
with altered nonperturbative sectors are the same as nonlinear sigma
models on special stacks known as gerbes.
After reviewing the
existing results on such theories in two dimensions, we discuss
examples of gerby moduli `spaces' appearing in four-dimensional field theory and
string compactifications, and the effect of various dualities.  
We discuss global topological defects arising when a field or string theory
moduli space has a gerbe structure.  We also outline how to
generalize results of
Bagger-Witten and more recent authors on quantization issues
in supergravities from smooth moduli spaces to smooth moduli stacks,
focusing particular attention on stacks that have gerbe structures.

\begin{flushleft}
December 2010
\end{flushleft}

\newpage

\tableofcontents

\newpage

\section{Introduction}

Recently there has been much progress in understanding
Fayet-Iliopoulos parameters in supergravity, generalizing
work of {\it e.g.} Bagger-Witten \cite{bag-ed1},
see for example
\cite{zohar1,zohar2,zohar3,butter1,zohar4,nati0,git-sugrav,banks-seib}.
In particular, the recent paper \cite{nati0} argued that in the special case of
linearly-realized group actions, Fayet-Iliopoulos parameters could be 
interpreted as charges for a $U(1)$ gauge symmetry, and so are
quantized.  This result was generalized in \cite{git-sugrav} to 
the more nearly generic case of
nonlinearly-realized group actions, by demonstrating that the Fayet-Iliopoulos
parameters determine the lift of the group action to the Bagger-Witten 
\cite{bag-ed1}
line bundle.  As such lifts of group actions are quantized, the
Fayet-Iliopoulos parameters are therefore also quantized.

This paper will focus on 
another aspect of \cite{nati0}, specifically, a proposal for Fayet-Iliopoulos 
quantization when the moduli space is defined by two-dimensional
sigma models with a restriction on
allowed instantons.  Such two-dimensional theories
have been discussed previously in {\it e.g.}
\cite{kps,nr,msx,glsm,hhpsa,cdhps,tonyme},
and are the same
as sigma models on gerbes, special kinds of stacks.

Schematically, smooth stacks are ``manifolds paired with automorphisms.'' 
Stacks 
admit metrics, spinors, and all the other structures appearing in
classical field theories.  The original interest in stacks in the physics
community revolved around using them to form new string compactifications,
new conformal field theories, and applying them to give a more fundamental
understanding of certain existing compactifications.

Previous work on consistency conditions in
supergravity theories has assumed that the moduli space is a smooth
manifold.  However,
in mathematics, moduli `spaces' are usually stacks,
and not manifolds, so to have
a broad understanding of classical consistency
conditions on supergravity theories, one must understand
cases in which the moduli `space' of the supergravity is a stack.
This paper is a step in a program of understanding consistency
conditions for such more general cases.

To be more specific,
in this paper we will discuss generalizations of consistency conditions
on supergravities from moduli spaces that are manifolds to moduli
`spaces' that are smooth Deligne-Mumford stacks, focusing particular
attention on stacks that are gerbes over manifolds.  That said, in typical
examples arising in string compactifications, the moduli stack has
singularities, so our generalization to stacks will still not describe
all cases pertinent to string compactifications, but is a step towards
a directly pertinent treatment.

We begin in section~\ref{2drev} by reviewing two-dimensional sigma models on
stacks, focusing in particular on gerbes over manifolds.  Two-dimensional
sigma models on gerbes
over manifolds look like sigma models on the underlying manifolds but with
a restriction on topological sectors.  These have been discussed in considerable
detail in both the mathematics and physics literature, as we review.

In section~\ref{4dphys} we discuss analogous four-dimensional
theories.  There are some significant differences between two-dimensional
and four-dimensional cases, including issues around presentation
dependence, and (on ${\bf R}^4$) a lack of nonperturbative sectors in
gerbe theories.  

In section~\ref{exs-gerbes} we discuss particular
examples of both field and string theories whose moduli `spaces' are
gerbes over manifolds.  In particular, previous work on gerbe structures
in supergravity moduli spaces \cite{nati0} did not give any 
examples of string compactifications
in which such structures would arise, which we remedy here.
We discuss the physical impact of such gerbe
structures, and also discuss the action of duality groups.

In section~\ref{top-defects} we discuss global topological defects in
theories with gerby moduli spaces.  Topological defects are classified
by homotopy of the moduli space, and gerbe structures contribute
nontrivially to the homotopy.  We discuss whether the contributions to
homotopy from gerbe structures have physical meaning.

In section~\ref{sugrav-stacks} we outline how to
generalize consistency conditions
on classical supergravities in \cite{bag-ed1,git-sugrav} to moduli `spaces'
that are smooth Deligne-Mumford stacks, 
focusing in particular on the case of stacks that are
gerbes.  In particular, we discuss the case of Bagger-Witten \cite{bag-ed1}
line bundles that are `fractional' over the gerby moduli space.

In appendix~\ref{4d-decomp} we discuss a four-dimensional
analogue of the `decomposition conjecture' \cite{hhpsa} that plays a 
vital role in understanding two-dimensional sigma models on gerbes.
In this four-dimensional analogue, we restrict sums over four-dimensional
instantons -- as a result, the four-dimensional version is not directly
relevant to four-dimensional sigma models on gerbes, but 
nevertheless we thought it
appropriate to discuss here.

Finally, in appendix~\ref{2dbf} we discuss two-dimensional $BF$ theory
and analogues of gerbe structures and decomposition statements there.
This gives us an opportunity to discuss the relationship between
locality and cluster decomposition in an explicit example.

While this work was being completed, the paper \cite{banks-seib}
appeared, which has nontrivial overlap.

\section{Review of two-dimensional theories with altered topological sectors}
\label{2drev}

The recent paper \cite{nati0} discussed theories defined by restricting
sums over instantons to a subset of all instantons.
In this section we briefly review some of the previous work
done on such theories.

In the case of two-dimensional nonlinear sigma models,
a nonlinear sigma model in which the sum over worldsheet instantons is
restricted to a subset of all instantons is the same as a string
on a gerbe, a special kind of stack, as is discussed in the physics
literature in for example
\cite{kps,nr,msx,glsm,hhpsa,cdhps,tonyme,karp1,karp2}
and reviewed in conference proceedings including 
\cite{me-vienna,me-tex,me-qts}.
(There is also a significant mathematics literature on Gromov-Witten
invariants of stacks and gerbes; see for example 
\cite{cr,agv,cclt,mann} for a few representative examples.)

Briefly, a stack is a manifold ``paired with automorphisms.''  
(See {\it e.g.} \cite{vistoli,gomez,lmb} for a more technical definition.)
At the same
level of brevity, a gerbe is a stack
in which one has the same automorphisms everywhere.  Mathematically,
a gerbe can be thought of locally as covered by patches of the form
$[U/G]$ where $U$ is an open set and $G$ acts trivially on $U$.
Stacks keep track of even trivial group actions, and so $[U/G]$ is
distinguished (as a stack) from just $U$.  

One of their properties that
plays a role in this paper is that if 
${\cal G}$ is a gerbe over a manifold $M$, then maps
into ${\cal G}$ are equivalent to maps into $M$ with a restriction on
their degree, as discussed in for example \cite{glsm}.
Briefly, a map from any space $X$ into a gerbe ${\cal G}$ over $M$
is equivalent\footnote{
There is a closely related statement for bundles.  Given a map $g:
X \rightarrow
E$ for some bundle $\pi: E \rightarrow M$, we can compose with $\pi$ to
produce a map $f: X \rightarrow M$.  Furthermore, we can define a 
trivialization of $f^* E$ canonically:  recall
\begin{displaymath}
f^* E \: \equiv \: \left\{ (x, e) \in X \times E \, | \,
f(x) \: = \: \pi(e) \right\}
\end{displaymath}
so we can define a trivialization $X \rightarrow f^* E$ by
$x \mapsto (x, g(x))$ for $x \in X$.  Conversely,
If $E$ is a bundle over
a space $M$, and we are given a map $f: X \rightarrow M$ and
a trivialization of $f^* E$, then since there
is a canonical map $f^* E \rightarrow E$, the trivialization
$X \rightarrow f^* E$
can be composed with that canonical map to form a map $g: X \rightarrow E$,
whose composition with the projection is $f$ again.   
}
to a map $f: X \rightarrow M$ into the underlying space, together with
a trivialization of $f^* {\cal G}$.  For a ${\bf Z}_k$ gerbe, say,
$f$ induces a map
\begin{displaymath}
f^*: \: H^2\left( M, {\bf Z}_k \right) \: \longrightarrow \:
H^2\left(X, {\bf Z}_k \right)
\end{displaymath}
which maps the characteristic class of the gerbe ${\cal G}$
(an element\footnote{
In general, the characteristic class of a $G$-gerbe on a manifold
$X$ is an element of
$H^2(X, C^{\infty}(G))$.  Here, since $G$ is assumed finite,
$C^{\infty}(G) = G$.  For $U(1)$ gerbes, 
\begin{displaymath}
H^2(X, C^{\infty}(U(1)) ) \: = \: H^3(X, {\bf Z})
\end{displaymath}
and in this fashion one recovers the usual physics description in terms of
the curvature $H$ of the $B$ field.
} in $H^2(M, {\bf Z}_k)$)
to the characteristic class of $f^* {\cal G}$, which should vanish 
(otherwise $f^* {\cal G}$ would not admit a trivialization).
For example, if $X = {\bf P}^1$ and $M = {\bf P}^{N-1}$,
then $f$ is characterized by an integer, its degree.  In this case,
$f^*$ acts by multiplication by the degree of $f$, so if we let $n$
denote the characteristic class of the gerbe (mod $k$), and $d$ the
degree of $f$, then we have the constraint
\begin{displaymath}
d \left(n \mbox{ mod }k \right) \: = \: 0 \mbox{ mod }k.
\end{displaymath}
In other words, $dn$ must be a multiple of $k$, a constraint on the
allowed degrees of maps $f$.
(Note the constraint depends upon the characteristic class of the
gerbe -- for example, for a trivial gerbe, $n \equiv 0 \mbox{ mod }k$,
and so there is no constraint.)

In mathematics, moduli `spaces' are usually stacks, hence one should not be
surprised to find stack structures arising in moduli spaces of interest
to physicists.  Indeed, in this paper we shall discuss examples of
moduli `spaces' with nontrivial stack structures arising in physics,
and their analysis.

In previous work {\it e.g.} \cite{nr,msx,glsm}, two-dimensional
sigma models on smooth
Deligne-Mumford stacks were defined physically by using the fact that
essentially\footnote{See \cite{msx} for a discussion of rare counterexamples,
and their apparent lack of physical relevance.}
all such stacks can be described as quotients $[X/G]$ for $X$ some space
and $G$ some group acting on $X$.  The special case of a gerbe
corresponds to 
a subgroup of $G$ acting trivially on $X$.
(We shall review how physics keeps track of even trivial group actions.) 
To such a quotient we associate a 
$G$-gauged sigma model on $X$.  A given stack can admit multiple presentations
of this form; we associate universality classes of renormalization group flow
to particular stacks.  Much effort was expended in previous work to
check presentation-independence of universality classes.

A standard example\footnote{
This ${\bf Z}_k$ gerbe has characteristic class $-1 \mbox{ mod }k$,
so from the previous analysis, maps into the gerbe are maps into the
underlying projective space of degree divisible by $k$, as should also 
be clear from the description of the gauge theory.
} of a gerbe is a ${\bf Z}_k$ gerbe on a projective
space, defined by a supersymmetric ${\bf P}^{n-1}$
model, a linear gauge theory, with $n$ chiral superfields
$\Phi_i$ each of charge $k$ instead of charge 1.
As discussed in \cite{nr}, in two dimensions
such theories are
nonperturbatively distinct from the ordinary ${\bf P}^{n-1}$ model.
On a noncompact worldsheet, this can be seen by thinking about periodicity
of the two-dimensional theta angle -- such theories can be distinguished
by the existence of massive minimally-charged objects, which alter the
periodicity. 

Let us work through this argument in more detail.
As described in \cite{nr}[section 3],
since in two dimensions the $\theta$ angle couples to $\mbox{Tr }F$,
we can determine the instanton numbers through the periodicity of
$\theta$.  Suppose we have the physical theory described above,
namely a GLSM with Higgs fields of charge $k$,
plus two more massive fields, of charges $+1$ and $-1$.
In a two-dimensional theory, the $\theta$ angle acts as an electric
field, which can be screened by pair production, and that screening
determines the periodicity of $\theta$.
If the only objects we could pair produce were the Higgs fields
of charge $k$, then the theta angle would have periodicity
$2 \pi k$, and so the instanton numbers would be multiples
of $1/k$.  However, since the space is noncompact, and the
electric field fills the entire space, we can also pair produce
arbitrary numbers of the massive fields, which have charges
$\pm 1$, and so the $\theta$ angle has periodicity $2 \pi$,
so the instantons have integral charges.
In particular, even if the masses of the massive fields are beyond the
cutoff scale, the theta angle periodicity can still know about them,
and so they can still help determine the low-energy effective field
theory.

We can phrase this more simply as follows.
In a theory with only Higgs fields of charge $k$,
the instanton numbers are multiples of $1/k$, and so the resulting
physics is equivalent to that of a GLSM with minimal charges.
However, if we add other fields of charge $\pm 1$,
then the instanton numbers are integral,
and if those fields become massive, and we work at an energy scale
below that of the masses of the fields, then we have a theory
with Higgs fields of charge $k$, and integral instanton numbers,
giving us the physics that corresponds to a gerbe target.
(This argument was, to our knowledge, first developed by
J.~Distler and R.~Plesser at an Aspen summer meeting in 2004,
used with their permission in \cite{nr}[section 3] and 
then also described much more recently in
\cite{nati0}.)

On a compact worldsheet, this distinction between minimal and nonminimal
charges is a consequence of how matter couplings
are defined globally ({\it i.e.} as sections of bundles).
In detail, to uniquely define the theory nonperturatively on
a compact space, we must specify, by hand, the bundles that the
Higgs fields couple to.  If the gauge field is described by
a line bundle $L$, then coupling all of the Higgs fields to
$L^{\otimes k}$ is a different prescription from coupling all
of the Higgs fields to $L$.  As a result, the spectrum of zero modes
differs between the two theories, hence correlation functions and
anomalies differ between the two theories.

Some of the structure above -- such as the theta angle argument -- is
specific to two dimensions, but some will generalize.  Later, we will argue
that gerbes are relevant to four-dimensional physics when either the
four-dimensional spacetime is topologically nontrivial, or there are
massive states which are not invariant, mirroring aspects of the
two-dimensional story above.

In any event, here are a few physical consequences of this
distinction between minimally and nonminimally-charged theories
in this two-dimensional example \cite{glsm}:
\begin{itemize}
\item The axial $U(1)_A$ of the supersymmetric ${\bf P}^{n-1}$ model
is broken to ${\bf Z}_{2kn}$ by instantons
instead of ${\bf Z}_{2n}$, when the chiral superfields 
have charge $k$.  
\item The quantum cohomology ring is ${\bf C}[x]/(x^{kn} - q)$
instead of ${\bf C}[x]/(x^n - q)$, reflecting the fact that A model
correlation functions differ.
(The paper \cite{glsm} provided GLSM-based
physical computations of these quantities, as well as a mirror symmetry
computation verification,
and in addition there are also mathematical definitions and matching results;
see for some representative examples \cite{agv,cr,cclt,mann}.)
\item The (Toda) Landau-Ginzburg mirror can be described by a 
superpotential with a field $\Upsilon$ valued in $k$th roots of
unity, 
\begin{displaymath}
W \: = \: e^{Y_1} \: + \: e^{Y_{n-1}} \: + \: \Upsilon e^{-Y_1 - \cdots -
Y_{n-1}},
\end{displaymath}
reflecting both the fact that the theory on a gerbe is equivalent to a theory
on a disjoint union of spaces, and the fact that mirror symmetry dualizes
nonperturbative effects into perturbative ones.
(This result was physically derived in \cite{glsm} from duality for GLSM's
ala \cite{hv,mp},
and also independently derived in {\it e.g.} \cite{mann} from mathematical
considerations.)
\end{itemize}

So far we have outlined how noneffective continuous group actions
can lead to new physics; the same is true of finite group actions.
For example, consider the orbifold $[T^6/D_4]$ where $D_4$ is
an eight-element group that projects onto ${\bf Z}_2 \times {\bf Z}_2$:
\begin{displaymath}
1 \: \longrightarrow \: {\bf Z}_2 \: \longrightarrow \: D_4 \:
\longrightarrow \: {\bf Z}_2 \times {\bf Z}_2 \: \longrightarrow \: 1.
\end{displaymath}
To specify the orbifold, we must specify the action of $D_4$ on $T^6$.
Let us take the ${\bf Z}_2$ center to act trivially, so that the $D_4$
acts by first projecting to ${\bf Z}_2 \times {\bf Z}_2$, and then act with
a standard Calabi-Yau action of  ${\bf Z}_2 \times {\bf Z}_2$ on
$T^6$, as described in {\it e.g.} \cite{vw-dt}.  Since the ${\bf Z}_2$
center acts trivially, one might naively assume that the $[T^6/D_4]$
orbifold would be physically equivalent to a $[T^6/{\bf Z}_2 \times {\bf Z}_2]$
orbifold.  Instead, one computes that at one-loop, for example,
\cite{hhpsa}[section 5.2]
\begin{displaymath}
Z\left( [T^6/D_4] \right) \: = \:
Z\left( [T^6/{\bf Z}_2 \times {\bf Z}_2] \coprod [T^6/{\bf Z}_2\times
{\bf Z}_2]_{\rm d.t.} \right),
\end{displaymath}
where the subscript indicates the presence of discrete torsion in one
of the two factors.  We therefore see explicitly that, in this example,
the string orbifold knows about the trivially-acting ${\bf Z}_2$
subgroups.  Many additional examples have been described
in detail in the references.
Thus, string orbifolds know about trivially-acting subgroups,
just as we saw in two-dimensional gauge theories.

Technically, in (2,2) supersymmetric cases,
these two-dimensional theories (in which a trivially-acting group have
been gauged) do not obey cluster decomposition.
(This is an immediate consequence of Weinberg's ancient argument for
theories with any restriction on instanton degrees,
and can also be seen for CFT cases by, for example, computing massless
spectra and observing multiple dimension zero operators.)
This would be a problem were it not for the fact that they are equivalent
to nonlinear sigma models on disjoint unions of ordinary spaces \cite{hhpsa},
a result described there as the `decomposition conjecture.'
The latter also do not obey cluster decomposition, but are obviously
nevertheless under good control, in the sense that we can renormalize
by local counterterms, and so forth.
Thus, they are sums of theories which obey cluster decomposition,
and so at least morally are ``local'' theories.

One of the original reasons for interest in these gerbe theories was
the idea that they might define new SCFT's, new string compactifications.
Because of the decomposition conjecture, that is not really the case
in (2,2) theories, as one gets sums of existing theories.
In (0,2) theories, on the other hand,
the story seems to be somewhat more complex;
an example is outlined in \cite{kap-02}[section 3.2], 
and a more complete description
will appear in \cite{tonyme}.

We can understand the decomposition conjecture schematically as follows.
Consider a nonlinear sigma model on a space $X$, for simplicity with
$H^2(X,{\bf Z}) = {\bf Z}$, with a restriction on
worldsheet instantons to degrees divisible by $k$.  We can realize that
restriction in the path integral by inserting a projection operator
\begin{displaymath}
\frac{1}{k} \sum_{n=0}^{k-1} \exp\left( i \int \phi^* \left( 
\frac{2 \pi n}{k} \omega \right) \right),
\end{displaymath}
where $\omega$ is the de Rham image of a generator of $H^2(X,{\bf Z})$.
Inserting this operator into a partition function is equivalent to working
with a sum of partition functions with rotating $B$ fields,
and this is the essence of the decomposition conjecture.

One of the applications of the result above is to Gromov-Witten theory,
where it has been checked and applied to simplify computations of
Gromov-Witten invariants of gerbes, see \cite{ajt1,ajt2,ajt3,t1,gt1,xt1}.
Another application is to gauged linear sigma models \cite{cdhps}, where it 
answers old questions about the meaning of the Landau-Ginzburg point
in a GLSM for a complete intersection of quadrics, as well as corrects
old lore on GLSM's.

\section{Four-dimensional physics}   \label{fourdim-basics}
\label{4dphys}

Now, let us turn to four-dimensional theories, and discuss how the physics
differs from two-dimensional cases.

To sharply compare with the two-dimensional cases outlined in the
previous section, let us work through a toy example.
Consider a $U(1)$ gauge theory in
supergravity\footnote{
The analysis presented here is due to J.~Distler, and we thank him for
allowing us to present it here.
}, defined over spacetime ${\bf R}^4$, 
with supergravity moduli space
${\bf C}^{2n+2}$, describing $2n+2$ complex scalars,
on which the $U(1)$ acts as follows:
$n$ fields of charge $k$, $n$ fields of charge $-k$,
one field $\chi$ of charge $+1$, one field $\tilde{\chi}$ of charge $-1$.
Let us furthermore
assume that there is a superpotential\footnote{
Since we are working in supergravity,
the superpotential is a section of the
Bagger-Witten \cite{bag-ed1} line bundle, 
but since the supergravity moduli space is
${\bf C}^{2n+2}$, the Bagger-Witten line bundle is necessarily
the trivial line bundle ${\cal O}$.
} $W = m \chi \tilde{\chi}$,
giving a mass $m$ to the two
fields of charge $\pm 1$.  
The upshot of this construction is that at low energies, one has a 
$U(1)$ gauge theory with nonminimal charges, closely analogous to those
discussed in the last section realizing sigma models on gerbes.

Now, one might worry that at at low energies, below the cutoff scale,
perhaps all the states
of the theory have charges that are a multiple of $k$.  In such a case,
the fact that the electron charges above are 
nonminimal would be physically irrelevant; at low energies, the theory would
be physically equivalent to a theory in which all fields had charge $1$,
not $k$.
To settle this issue, we
need to understand the correct electric charge quantization in this theory.
In two dimensions, we could distinguish a theory with nonminimal charges
from a theory with minimal charges via nonperturbative effects, invoking
the theta angle periodicity to `see' states with masses beyond the cutoff.
Here,
by contrast, note that since we are describing a $U(1)$ gauge theory on
${\bf R}^4$, there are no $U(1)$ instantons.
However, because this theory is coupled to gravity, we can appeal to the
existence of Reissner-Nordstr\"om black holes.  We can use them to determine
the correct electric charge quantization in
the theory at places in its moduli space
where the $U(1)$ is unbroken, and then appeal to continuity to understand
the rest.

First, suppose that the mass $m < M_{\rm Pl}$.
When the $U(1)$ is unbroken,
there are electrically-charged Reissner-Nordstr\"om black holes.
Since $m < M_{\rm Pl}$, microscopic black holes can Hawking radiate
$\chi$, $\tilde{\chi}$, and so even if one started with a black hole
of charge a multiple of $k$, it could Hawking radiate down to charge $1$.
Thus, at least at points where the $U(1)$ is unbroken, the nonminimal
charges of the Higgs fields are physically relevant.
At more generic points on the moduli space, where
the $U(1)$ is Higgsed to ${\bf Z}_k$, we need there to be excitations on which
the ${\bf Z}_k$ acts nontrivially, and at least for small Higgs vev,
the Reissner-Nordstr\"om black holes should\footnote{
In effect, we are appealing to a continuity argument.  As a potential
loophole, we should mention that it is known
from work on wall-crossing that the spectrum of BPS states in a supersymmetric
theory does not always behave so simply.  We do not need to assume the
states here are BPS, but, it is possible that analogous processes may
apply.
} become such excitations.

If $m > M_{\rm Pl}$, then the Hawking radiation process above can not happen,
but demanding that physics be continuous in $m$ leads us to believe that the
electric charges of black holes are still multiples of $1$ rather than $k$.
Thus, again, the fact that the massless fields in the gauge theory have
nonminimal charges, is physically relevant.

So far we have discussed a four-dimensional analogue of the two-dimensional
theta-angle-periodicity argument for the relevance of nonminimal charges,
using black holes rather than theta angles.  In the rest of this section
we shall discuss some subtleties specific to four-dimensional theories,
and their relevance to four-dimensional sigma models on stacks.

First, let us examine more closely the 
(lack of) nonperturbative sectors in field
theories on ${\bf R}^4$ not coupled to gravity.
One of the significant properties of two-dimensional
sigma models on gerbes was that the nonperturbative sector is altered:
one sums over only some instantons, not all of them.
However, in four dimensions, on ${\bf R}^4$, there is no constraint on
nonperturbative sectors:
\begin{itemize}
\item We have already seen the example of
a $U(1)$ gauge theory with nonminimal charges on
${\bf R}^4$.  
Instantons in such a theory
would be described by principal $U(1)$ bundles on $S^4$
(the one-point compactification of ${\bf R}^4$, taken to encode
compact-support issues), and there are no nontrivial principal
$U(1)$ bundles on $S^4$, because $H^2(S^4, {\bf Z})$ vanishes.
(Moreover, on ${\bf R}^4$ or $S^4$, for a $U(1)$ gauge theory the
quantity $\int {\rm Tr} F \wedge F$ vanishes identically for all gauge field
configurations appearing in the path integral ({\it i.e.}
compactly-supported ones), not just saddle points.)
\item Next, formally consider a nonlinear sigma model summing over
maps ${\bf R}^4 \rightarrow
{\cal G}$ for ${\cal G}$ some $G$-gerbe (for finite $G$) over a smooth
manifold $M$.  Again for compact-support reasons we can replace ${\bf R}^4$
by $S^4$ for the purposes of mathematical computations.
As discussed in the last section, a map $X \rightarrow {\cal G}$ for
any space $X$ is the same as a map $f: X \rightarrow M$ together with a
trivialization of $f^* {\cal G}$, and demanding that $f^* {\cal G}$
be trivializable restricts possible maps $f$.  In the present case, however,
since $H^2(S^4, {\bf Z})$ vanishes, the pullback
$f^* {\cal G}$ is always trivializable for any ${\cal G}$,
and so demanding that
$f^* {\cal G}$ be trivial is no longer a constraint on possible maps.
\end{itemize}
In particular, unlike two dimensions, on ${\bf R}^4$ there is no reason
to believe that cluster decomposition will necessarily be violated,
and there is no analogue of the decomposition conjecture
\cite{hhpsa} for gerbe theories.

Next, let us
consider a technical point in
the renormalization-group flow of the
low-energy effective gauged sigma models arising in this and
analogous examples.  (We would like to thank J.~Distler for patient
explanations of this point.)
Schematically, if $v$ is the scale of the Higgs vev, and $g$ the coupling,
then the low-energy effective action is an expansion in powers of $E/v$.
However, Higgsed gauge fields have masses which scale as $g v$, and so for
weak coupling $g$, generate $E/(gv)$ effects which can be 
stronger than low-order
effective action terms.  Put another way, W bosons are light relative to the
natural mass scale defining the metric.
As a result, the effective field theory arising 
in the infrared from a gauged sigma model often can not be the same as a 
nonlinear sigma model.  There can still be a moduli space, a metric
on the moduli space, and many other features consistent with nonlinear
sigma models, (as happens with {\it e.g.} Narain moduli spaces in
toroidally-compactified heterotic strings,) but strictly speaking,  
the infrared limit of a low-energy effective gauged sigma model 
in four dimensions need not be physically equivalent to a nonlinear
sigma model.

This result implies an issue of presentation-dependence in four-dimensional 
theories, that does not exist in two dimensions.
In two dimensions, we identify universality classes of renormalization
group flow with stacks:  a given stack can have multiple presentations
with different UV physical descriptions (a nonlinear sigma model,
a gauged sigma model, an orbifold) which mathematically correspond to
the same stack.  Physically, it is conjectured that those different
presentations lie in the same universality class, that the
renormalization group `washes out' all details of the presentation,
leaving physics that only depends upon the stack and not how it is
described or presented.
In particular, typically we are only interested in conformal field theories
arising at endpoints of renormalization group flow, so the details of a
physical presentation of a massive UV theory are irrelevant.

Part of the point of the observation on four-dimensional low-energy
effective field theories is that the presentation-independence
we enjoyed in two dimensions no longer applies in four dimensions.
We can also see this from another perspective, involving the
gauge kinetic terms.  A sum over maps from a space into a stack
presented as
$[X/G]$ involves a sum over $G$ bundles with connection -- a sum over
$G$-gauge fields.  In two dimensions, gauge kinetic terms are generated
dynamically, so we could effectively ignore them and identify
a nonlinear sigma model on $[X/G]$ with a $G$-gauged sigma model on $X$
 -- the sum over maps includes the sum over gauge fields, and the
gauge kinetic term comes for free.  By contrast, in four
dimensions, gauge kinetic terms are not generated dynamically.
We can describe some aspects of a $G$-gauged sigma model on $X$ with
the stack $[X/G]$, but we do not get a gauge kinetic term automatically,
the stack does not even determine a classical value of the gauge
coupling.  Hence, in four dimensions, merely specifying a stack
does not uniquely determine the physics.

In two dimensional cases, we associated stacks with universality classes
of renormalization group flow.  
Instead, in four dimensions we will
use stacks as `universal' objects from which various different physical
presentations can be associated.  The details of those presentations,
the presentation-dependence, will no longer be physically irrelevant,
unlike two dimensions;
nevertheless, some (not all)
of the physics will be determined by the stack.
It is in this sense that we will associate stacks  
with low-energy effective four-dimensional (gauged) sigma models and
related theories.

Before going on, let us summarize the circumstances under which 
gerbes will be physically meaningful in a four-dimensional theory.
One way for gerbes to be physically meaningful in a four-dimensional theory
is if the four-dimensional spacetime is topologically nontrivial,
with nonzero $H^2({\bf Z})$.  In this case,
one would have nontrivial nonperturbative
sectors in the examples above.  Just as in two dimensions, to uniquely
define Higgs fields one would need to specify the precise bundle the field
couples to, and that choice would be reflected in zero mode spectra, hence
in anomalies and so forth.  
A second way gerbes can be physically meaningful is if there
are massive non-invariant states.  We have only argued this above in
theories coupled to gravity; however, we shall also see examples later
where some aspects of gerbes manifest even in four-dimensional
theories on ${\bf R}^4$ not coupled to gravity.
Both of these cases had analogues
in two dimensions -- for example, the theta-angle-periodicity argument in
two dimensions relied on the existence of massive minimally-charged states.
We shall see examples of both cases in the next section.

Partly with an eye towards nonabelian applications we shall discuss later
in this paper,
let us summarize the conclusions of this and the previous section in the
following slogan:
\begin{quotation}
Perturbative physics is determined by the Lie {\it algebra} of the gauge
group, but nonperturbative physics is determined by the Lie {\it group},
not just the algebra.
\end{quotation}
Just as an asymptotic series expansion does not
uniquely determine the function it is expanding, so too does the perturbative
physics not uniquely determine the nonperturbative physics.

In the rest of this paper, we will outline gerbes in four dimensional
field theories and string compactifications.

\section{Examples, duality in gerby moduli `spaces'}
\label{exs-gerbes}

Gerby moduli 'spaces'\footnote{
Strictly speaking, if there is a gerbe structure, then the moduli
`space' is actually a stack, not a space, but because the language of
stacks is as yet unfamiliar to many physicists, we will call
them ``gerby spaces'' in much of this paper.}
seem to appear in both four dimensional field theory and in string
compactifications, as we shall outline in this section.
Briefly, a gerbe looks locally like a quotient by a trivially-acting
group -- although the group acts trivially, both mathematics and,
at least sometimes, physics nevertheless knows about the group action.
As sigma models on gerbes can be viewed as sigma models on spaces or
effective quotients with a restriction on nonperturbative sectors,
these are precisely the examples discussed recently in \cite{nati0}.

In this section we shall discuss examples of gerby moduli spaces appearing
in both field and string theories, and also discuss how the gerbiness
behaves under field and string theoretic dualities.

\subsection{Field theory}

At a purely mathematical level, 
it is easy to generate examples of four dimensional
field theories with gerbe structures over their moduli spaces.
As the moduli space of a field theory is typically of the 
form $[V/G]$, where $V$ is a vector space spanned by matter vevs
and $G$ is the gauge group, whenever any subgroup of $G$ acts
trivially on all of the massless matter, mathematically one could associate
a gerbe structure to that moduli space.  
For example, in Yang-Mills theory with adjoint matter,
the maximal torus of the gauge group acts trivially on matter vevs.
Thus, if $r$ is the rank of the gauge group, then in such theories
there is formally a $U(1)^r$ gerbe structure generically\footnote{
This stabilizer changes over the moduli space; for example, at the origin
where all vevs vanish, the stabilizer is all of $G$.  Since the
stabilizer changes, this is not, strictly speaking, a gerbe, but rather
is a more general stack, that only looks like various gerbes on specific
strata
}.  
The physical content of that gerbe structure is another matter.
Morally, if not literally\footnote{
A stack with non-finite stabilizers is known as an Artin stack.
The geometric interpretation of Artin stacks is somewhat more complicated
than that of Deligne-Mumford stacks, which the analysis of
\cite{kps,nr,msx,glsm,hhpsa,cdhps,me-vienna,me-tex,me-qts,tonyme} focused on.
In this paper we also almost exclusively focus on Deligne-Mumford stacks.
}, a sigma model on $U(1)^r$ gerbe ought to be a $U(1)^r$ gauge theory,
which certainly arise in Yang-Mills theories with only adjoint matter.

However, we need a bit more structure (such as massive
noninvariant matter, or a topologically-nontrivial spacetime
four-manifold) before we believe such gerbe structures are
physically meaningful.  In addition, in this paper we will focus
on finite gerbe structures (corresponding to Deligne-Mumford stacks,
rather than Artin stacks).
In the examples we shall discuss in this section, the gerbe
structure will arise by
focusing on the center of the gauge group.
If we return again to Yang-Mills theories with adjoint matter,
this means we consider the gerbe structure on the moduli space arising from
the fact that the center acts trivially on the matter.

Our first physical example will involve a topologically-nontrivial
spacetime four-manifold.
Consider 
${\cal N}=4$ supersymmetric theories arising
in recent work on the geometric Langlands program \cite{ed-anton}.
There, one compactifies a four-dimensional 
${\cal N}=4$ theory along a Riemann surface to get a two-dimensional
theory, a nonlinear sigma model whose target space is the Hitchin moduli space
on the compactification curve.  
The authors of \cite{ed-anton} observed that said moduli space has a
number of components.  An alternative way of understanding that fact is to
utilize the finite gerbe story outlined above.  If we start with a 
$G$ gauge theory in four dimensions, then following the ansatz above,
the moduli space of the four-dimensional theory (and hence the target
of the compactified two-dimensional sigma model) has a $Z(G)$ gerbe
structure, where $Z(G)$ denotes the center of $G$.
Application of the decomposition conjecture of \cite{hhpsa} to 
the two-dimensional sigma model on the gerbe then quickly reproduces
the multiple component structure worked out more painfully by
\cite{ed-anton}, as discussed in \cite{hhpsa,edgl2}.

One lesson of the example from geometric Langlands above is that these
formal gerbe structures on moduli spaces do have physical content -- 
the disconnectedness of the target of the two-dimensional sigma model is
a consequence of a gerbe structure on the moduli space.
That said, duality often does not preserve centers of gauge groups:
for example, S-duality in ${\cal N}=4$ maps $SU(n)$ gauge theories
to $SU(n)/{\bf Z}_n$ gauge theories.  In effect, the center of the gauge
group is being exchanged for extra characteristic classes, disconnectedness
in the two-dimensional target moduli space.  Hence, gerbe structures are not
duality-invariant.

We are often used to moduli spaces being invariant under duality operations
-- this is, after all, one of the standard checks of a duality.
What is going on here is that the underlying space is unchanged, only the
automorphisms that are paired with the space are changing.  Therefore,
the number of flat directions, the geometry of the flat directions is
unchanged, only the automorphisms differ.  Since it is only the
number and geometry of the flat directions that must necessarily
be preserved by duality, the fact that gerbe structures change does not
contradict duality.

Let us examine this ${\cal N}=4$ duality in greater generality.
Geometric Langlands exchanges the center $Z(G)$ with
the dual of $\pi_1({}^LG)$, where ${}^LG$ denotes the Langlands dual to
$G$.  The center $Z(G)$ encodes a gerbe structure, and $\pi_1({}^LG)$
describes how the moduli space breaks into components
(indexed by a characteristic class in $H^2(X,\pi_1({}^LG))$).
We can see how $Z(G)$ and $\pi_1({}^LG)^*$ are exchanged as follows.
Let $M$ denote the weight lattice of the Lie group $G$.
It is a sublattice of the weight lattice of the corresponding
Lie algebra, which we shall denote $\Lambda$.  ($M$ is determined by
the representations of the Lie group, instead of the Lie algebra.)
If we let $R$ denote the root lattice, then in general
\begin{equation}    \label{3lattices-a}
R \: \subseteq \: M \: \subseteq \: \Lambda.
\end{equation}
The action of Langlands duality is to dualize each of these three
lattices:
\begin{eqnarray*}
R & \mapsto & {}^LR \: \equiv \: {\rm Hom}(\Lambda, {\bf Z}), \\
M & \mapsto & {}^LM \: \equiv \: {\rm Hom}(M, {\bf Z}), \\
\Lambda & \mapsto & {}^L\Lambda \: \equiv \: {\rm Hom}(R, {\bf Z}),
\end{eqnarray*}
and it is straightforward to see from~(\ref{3lattices-a}) that
\begin{displaymath}
{}^LR \: \subseteq \: {}^LM \: \subseteq \: {}^L\Lambda.
\end{displaymath}
In this language, the center and $\pi_1$ of $G$ are determined by the
lattices above, as follows:
\begin{eqnarray*}
Z(G) & = & \left( M / R \right)^* \: = \: {\rm Hom}\left( M / R, {\bf Z}
\right),\\
\pi_1(G) & = & \left( \Lambda / M \right)^* 
\: = \: {\rm Hom}\left( \Lambda / M, {\bf Z} \right)
\end{eqnarray*}
(the first equality comes from the fact that $Z(G)$ is the kernel of
the adjoint action, whose weights generate the root lattice),
which should make it clear that
\begin{displaymath}
Z(G) \: = \: \pi_1({}^LG)^*, \: \: \:
\pi_1(G) \: = \: Z({}^LG)^*.
\end{displaymath}
In other words, Langlands duality exchanges the center of a group $G$ with
(the dual of) $\pi_1$ of the Langlands dual group ${}^LG$.

The Hitchin moduli stack, the target of the two-dimensional sigma model,
is a $Z(G)$-gerbe over a disconnected
space with multiple components.  One has different components
corresponding to the fact that there is a characteristic
class in $H^2(X, \pi_1(G))$, and the components are indexed by the value
of that characteristic class.  The effect of Langlands
duality is to exchange $Z(G)$ gerbiness with $\pi_1({}^LG)$ disconnectness
(see {\it e.g.} \cite{ron-tony} for a more detailed discussion).
One might ask if there is an alternative description as some
$Z(G) \times Z({}^LG)$ gerbe over another space, giving a 
duality-invariant stack, but we are told
\cite{tonypriv} such a construction does not exist.

Let us next consider some examples of gerbe structures appearing in the field
theories 
discussed in \cite{pouliot,poul-strass,strassler,strassler2}.
These papers discuss examples in which an ${\cal N}=1$
supersymmetric gauge theory with a gerbe structure on its moduli space
is (Seiberg-)dual to another ${\cal N}=1$ supersymmetric gauge theory
which has monopoles.  The massive, non-invariant matter on the gerbe side
is dual to the monopoles.
Just as in the geometric Langlands story above, the gerbe structure is
not preserved by duality.

The prototype for these examples is discussed in
\cite{poul-strass}.  
That paper argued that a Spin(8) gauge theory with $N_f$ fields in the
${\bf 8}_V$ and one field in the ${\bf 8}_S$ is dual to a chiral $SU(N_f-4)$
theory with a symmetric tensor and $N_f$ fields in the antifundamental
representation.  When the ${\bf 8}_S$ is given a mass, the dual
$SU(N_f-4)$ theory is Higgsed to $SO(N_f-4)$ with $N_f$ vectors.  
Moreover, that $SO(N_f-4)$ theory admits a monopole,
since $\pi_2( SU(N_f-4)/SO(N_f-4) ) = \pi_1(SO(N_f-4)) = {\bf Z}_2$.
The perturbatively massive spinor in the Spin(8) theory is dual to the
monopole in the $SO(N_f-4)$ theory.  

In the original (unHiggsed) dual pair, on neither side does the moduli space
admit a gerbe structure:  no part of the center of Spin(8) acts trivially
on both ${\bf 8}_V$ and ${\bf 8}_S$, and the center of $SU(N_f-4)$ does
not act trivially on the antifundamentals. 
After Higgsing, a ${\bf Z}_2$ subgroup of the center of Spin(8) acts
trivially on the remaining ${\bf 8}_V$ fields, hence that branch of the
moduli space (formally) admits a ${\bf Z}_2$ gerbe structure.
(Its dual still does not have a gerbe structure on its moduli space.)

The upshot is that we have two dual theories, one with a gerbe structure
on its moduli space and a massive spinor, dual to a theory without a 
gerbe structure on its moduli space, but with a massive monopole instead.
For example, a Wilson loop in the spinor representation of Spin(8) is
mapped to the 't Hooft loop in the magnetic ${\bf Z}_2$
\cite{strassler}.

Just as in geometric Langlands, we see that gerbe structures are not
preserved by duality.
This interpretation is reiterated (albeit without explicitly naming a gerbe
structure) in \cite{strassler}[section 2], \cite{strassler2}
in terms of screening effects,
and further examples of the same general form are given
in \cite{strassler,strassler2}.

For completeness, note that the presence of massive nonminimally charged
matter plays an important role in this story, just as it did in two-dimensional
examples of theories with gerbe structures.

\subsection{String theory}

Just as in field theory, one can also (formally) associate gerbe
structures to various moduli spaces, whenever there is a subset of the
low-energy gauge group that acts trivially on massless fields (and 
nontrivially on at least one massive field).  
In this section,
we will outline examples
of gerby moduli spaces appearing in string compactifications.

The first example is the Narain moduli space of toroidally-compactified
heterotic string theories.  Just as with Yang-Mills theories with
adjoint matter, there are at least two natural ways to formally
add a stack structure to such moduli spaces, both of which 
revolve around the fact that 
part of the Narain moduli space describes flat connections on a torus.
If we take the moduli stack of flat $G$-connections to be
\begin{displaymath}
\left[ {\rm Hom}(\pi_1, G) / G \right]
\end{displaymath}
then we have a stack which along strata has variable gerbe structures
along strata
(though as the stabilizer varies across strata, it is not considered
globally to be a gerbe, unless $G$ has a nontrivial center).  
For example, at the point on the moduli space
where low-energy adjoint scalars vanish, one has a $G$ gerbe;
at more nearly generic points, where only a maximal torus $T$ commutes
with adjoint scalars, one has a $T$ gerbe.  The mathematical
interpretation of such
structures is just as in the field theory discussion previously.

In the case of field theories, we observed that a different stack
structure may have greater physical relevance, involving only finite
centers of stabilizers rather than the entire stabilizer.  This structure
also varies across strata, giving rise along any one stratum to
a variety of possible gerbe structures.  Globally, the entire
stack would have a $Z(G)$ gerbe structure, 
where $Z(G)$ is
the center of $G$, which can be enhanced over various strata.  
In the case of geometric Langlands, this was the
gerbe structure that gave rise to the disconnectedness of the
target Hitchin moduli space of the two-dimensional theory.

Phrased more simply, ordinarily we think of toroidally-compactified
heterotic strings as having a Narain moduli space (or rather,
more generally \cite{triples}, 
a moduli space with several components, one of which
is the Narain moduli space).  Here we are observing that the
Narain moduli space (and other components) carry additional structure,
at least formally, namely that of a gerbe.  The moduli {\it stack} of
toroidally-compactified heterotic strings is a gerbe over a stack
with, in general, several components, one of which is the Narain
moduli space, plus enhanced gerbe structures on various strata.
(A more formal discussion of such phenomena in the context of
Hitchin moduli spaces can be found in {\it e.g.} \cite{ron-tony}.)

In the case of a Spin$(32)/{\bf Z}_2$ heterotic string compactification,
the moduli space of the toroidally-compactified
string theory generically (and formally) has
a ${\bf Z}_2$ gerbe structure, since the center of Spin$(32)/{\bf Z}_2$
is ${\bf Z}_2$.  As described elsewhere, for such a gerbe structure
to be meaningful for a theory on ${\bf R}^4$, 
we also need massive states which are not invariant
under the group.  In the present case, the Spin$(32)/{\bf Z}_2$ heterotic
string
has at its first massive level states transforming in the
spinor representation of Spin$(32)/{\bf Z}_2$, which is not invariant
under the ${\bf Z}_2$ (\cite{gsw1}[section 6.3.1], \cite{ghmr1}[section 2.3]), 
exactly as needed for a gerbe description of the moduli space
to be physically relevant.

On the ten-dimensional heterotic string worldsheet, 
this proposed ${\bf Z}_2$ gerbe structure on the
CFT moduli space manifests itself as the quantum symmetry\footnote{
The symmetry we are describing leaves the NS sector states invariant,
but multiplies the R sector states by a sign.  A ${\bf Z}_2$ quantum
symmetry leaves the untwisted sector invariant, and multiplies the twisted
sector by a sign, which is consistent with the symmetry in this case if
one remembers that we are using R, NS to describe states on the cylinder,
but the quantum symmetry is defined by 
(un)twisted sectors on the complex plane, and the conformal
transformation between the two exchanges R and NS sectors 
\cite{ginsparg}[section 7.1].
} 
\cite{vafa-qs} associated with the left-moving GSO analogue that
defines the Spin$(32)/{\bf Z}_2$ string in its RNS presentation.
(The center of Spin$(32)$ is ${\bf Z}_2 \times {\bf Z}_2$,
and the GSO analogue itself is responsible for the ${\bf Z}_2$ quotient in
Spin$(32)/{\bf Z}_2$.  As the action of the center
is being expressed on associated
vectors, not on the group itself, it manifests in terms of orbifolds and
quantum symmetries.  There is a closely analogous story for the
Spin$(16)/{\bf Z}_2$ in $E_8$ and RNS constructions of
$E_8 \times E_8$ heterotic strings.)
In particular, all of the massless ten-dimensional states arise from
a left-moving NS sector, and the only charged states are adjoints; 
all of the left-moving R sector states are massive.
The quantum symmetry leaves the left-moving NS sector states invariant
and multiplies the left-moving R sector states by a phase, which matches
the effect of the ${\bf Z}_2$ gerbe structure.

For completeness, let us also consider the string-dual type I theory in
ten dimensions.  The gauge group of the type I string is
$SO(32)$, different from that of its dual ten-dimensional heterotic
string.  We have seen that under dualities, gauge groups will change -- this
is a typical property of Langlands duality, for example.
The massless spectrum is invariant under the ${\bf Z}_2$ center of
$SO(32)$, suggesting a gerbe structure; however, all of the massive
states are also invariant, 
as the perturbative spectrum of the type I string in
ten dimensions contains only
symmetric and antisymmetric 2-tensors
\cite{ghmr1}[section 2.3], compatible in principle with
a gauge group
\begin{displaymath}
SO(32)/{\bf Z}_2 \: = \: {\rm Spin}(32)/\left( {\bf Z}_2 \times {\bf Z}_2
\right).
\end{displaymath}
For this reason, we do not identify a gerbe structure on the moduli space
of compactified type I strings.
Furthermore, in close analogy with our discussion of {\it e.g.} 
\cite{pouliot,poul-strass,strassler,strassler2} in the previous subsection,
there exists a particle-type topological defect in type I string theory,
arising from an element of $\pi_8(SO(32))$, which transforms as a spinor
of the Lie algebra \cite{ed-k}, 
the same property as a massive perturbative state in the dual heterotic 
Spin$(32)/{\bf Z}_2$ theory.

Along special loci this gerbe structure can be enhanced,
as expected on general grounds from our discussion of moduli stacks of
flat connections.
Consider Higgsing a toroidally-compactified $E_8 \times E_8$ heterotic
string, for example.  There is no gerbe structure over the entire
moduli space (as $E_8$ has no center).
Now Higgs one of the $E_8$'s to a Spin$(16)/{\bf Z}_2$
subgroup.  As all of the adjoint-valued scalars in the theory are derived
by dimensionally-reducing a ten-dimensional gauge field,
Higgsing the $E_8$ should lift all components of those scalars that
are not adjoints under Spin$(16)/{\bf Z}_2$.
The center of Spin$(16)/{\bf Z}_2$ is ${\bf Z}_2$,
and it acts trivially on the adjoints, the surviving 
massless matter.
However, it does not act trivially on all of the string modes.
The adjoint representation of $E_8$ decomposes as
\begin{displaymath}
{\bf 248} \: = \: {\bf 120} \: + \: {\bf 128}
\end{displaymath}
where ${\bf 120}$ is the adjoint representation of Spin$(16)/{\bf Z}_2$,
and ${\bf 128}$ is a spinor.  By Higgsing the $E_8$ to Spin$(16)/{\bf Z}_2$,
we give a mass to the ${\bf 128}$, which is not invariant under the
center of Spin$(16)/{\bf Z}_2$.  
Thus, we have, at low energies,
a gauge group with nontrivial (${\bf Z}_2$) center that acts
trivially on massless matter, but nontrivially on massive matter.

Note that we can construct 
examples with ${\cal N}-1$ supersymmetry in four dimensions and 
gerbe structures on their moduli spaces by compactifying a Spin$(32)/{\bf Z}_2$
heterotic string on a nontrivial Calabi-Yau threefold.
For simplicity, let us consider such a heterotic string compactification
with the standard embedding.

One way to see the existence of the gerbe structure on the moduli space
is from worldsheet considerations.  Just as in the ten-dimensional theory,
all of the massless states arise from left-moving NS sectors; the
left-moving R sectors contribute only massive states.  As a result,
the quantum symmetry associated to the left-moving GSO analogue
(which leaves left NS sectors invariant, but acts by a phase on
left R sectors) 
leaves the massless states invariant, but acts by a phase on 
massive states.

We can also see the gerbe structure on the moduli space in the
low-energy effective field theory.  Consider for simplicity a 
Spin$(32)/{\bf Z}_2$ heterotic
string compactification on a nontrivial Calabi-Yau threefold with the
standard embedding.  The low-energy gauge group is\footnote{
We can compute this as follows.  We are embedding $SU(3)$ into a
Spin$(6)=SU(4)$ subgroup, so we begin by observing that Spin(32) has
the subgroup
\cite{me-dist}[appendix A]
\begin{displaymath}
\frac{
{\rm Spin}(26) \times {\rm Spin}(6)
}{
{\bf Z}_2
}.
\end{displaymath}
Since the center of both Spin$(26)$ and Spin$(6)$ is ${\bf Z}_4$,
there is only one diagonally-acting ${\bf Z}_2$ subgroup.
We can describe the center of the group above as generated by $a$, $b$,
subject to the relations $a^4 = b^4 = 1$, $a^2 = b^2$.
Now, we want the subgroup of Spin$(32)/{\bf Z}_2$, and after taking the
second ${\bf Z}_2$ quotient we could have either a ${\bf Z}_2 \times
{\bf Z}_2$ or ${\bf Z}_4$ quotient of Spin$(26) \times$ Spin$(6)$, 
corresponding to quotienting either
$a^2$ or $ab$, respectively.  We can distinguish them as follows.
For simplicity, replace Spin$(26)$ by Spin$(6)$, to form subgroups of
Spin$(12)$, and use the fact that
Spin$(6) = SU(4)$, the ${\bf 4}$, ${\rm Alt}^3 {\bf 4} = {\bf \overline{4}}$
are the spinor representations, and ${\rm Alt}^2 {\bf 4} = {\bf 6}$ the
vector.  The ${\bf Z}_2$ quotient (originally of Spin$(32)$, now Spin$(12)$) 
should
flip the sign of the ${\bf 12} = ({\bf 6},{\bf 1}) \oplus ({\bf 1},{\bf 6})$,
and preserve only one of the two spinor representations.
Since $a$, $b$ both act by multiplying the ${\bf 4}$ by a fourth root of
unity, it is straightforward to check that $a^2$ leaves the vector 
representation
invariant, whereas $ab$ flips the sign of the vector representation.
(Both preserve only one spinor of Spin$(12)$.)  Thus, we should quotient by
$ab$, and hence the correct subgroup of Spin$(32)/{\bf Z}_2$ is
\begin{displaymath}
\frac{
{\rm Spin}(26) \times {\rm Spin}(6)
}{
{\bf Z}_4
}.
\end{displaymath} 
Embedding $SU(3)$ into $SU(4) = {\rm Spin}(6)$ leaves us with the
maximal commutant shown.
We would like to thank A.~Knutson for a useful discussion of this issue.
}
\begin{displaymath}
\frac{
{\rm Spin}(26) \times U(1)
}{
{\bf Z}_4
}.
\end{displaymath}
The $U(1)$ factor is typically anomalous and Higgsed via a four-dimensional
version of the Green-Schwarz mechanism \cite{dsw,ads}, closely related
to a (field-dependent, hence not directly relevant to this paper)
Fayet-Iliopoulos parameter.  The remaining ${\bf Z}_2$ center of
Spin$(32)/{\bf Z}_2$ descends to part of the center of the group
above, and the massless states are all invariant under this
${\bf Z}_2$, as the massless states all descend from invariant representations
of Spin$(32)/{\bf Z}_2$.

In the context of heterotic compactifications on elliptically-fibered
Calabi-Yau's, the moduli stack of $G$-bundles for a group $G$ has,
in essence, directions corresponding to moduli of the spectral cover
and directions corresponding to the moduli of a line bundle on the
spectral cover.  The latter has, at least formally, a $Z(G)$-gerbe
structure.  In the dual F theory compactification, such moduli dualize
to moduli of $G$ flux, suggesting \cite{tonypriv} that the moduli of
$G$ fluxes has a gerbe structure.  
In fact, naively not only do moduli spaces of F theory compactifications
admit gerbe structure, but at least sometimes there is evidence that
duals to some heterotic string compactifications are F theory compactifications
on gerbes \cite{tonypriv}.  Specifically, it has been observed
\cite{tonypriv} that the multisection structures appearing in
\cite{bps} in the F theory duals of heterotic CHL strings have an
alternative interpretation in terms of elliptic fibrations over ${\bf Z}_2$
gerbes.  Part of the point is that a heterotic compactification on an 
elliptic fibration with multisection is described by a
spectral cover in a gerbe over the relative Jacobian, together with 
a (possibly fractional, in a sense we describe later) line bundle 
over the restriction of the gerbe to the spectral cover.
The $G$ fluxes then behave as a torsor under the appropriate
Deligne cohomology group.
We shall not pursue such F theory structures further here.

We do not expect such gerbe structures to always appear in CFT
moduli spaces.  For example, consider a heterotic $E_8 \times E_8$ string
compactified on a simply-connected Calabi-Yau threefold, with the
standard embedding.  Although one has embedded an $SU(3)$ bundle,
and $SU(3)$ has a nontrivial center (${\bf Z}_3$), 
it has been embedded into an $E_8$,
which has no center.  We can determine the existence
of a gerbe structure by looking at the charged matter in the
low-energy effective field theory,
To get a gerbe 
structure, the matter would all have to be invariant under the center of the
low-energy effective gauge group (in this case, $E_6$, with center ${\bf Z}_3$).
Neither ${\bf 27}$'s nor ${\bf \overline{27}}$'s are invariant under
the center of $E_6$ \cite{allenpriv}, 
hence we do not expect to get a gerbe structure
on the CFT moduli space, since there is not a gerbe structure on the
field theory moduli space.

More generally, it is worth emphasizing that
many moduli spaces which do not have gerbe structures
globally will still have gerbe structures on subvarieties.  For a simple
example, the quotient stack $[{\bf C}^2/{\bf Z}_2]$ looks like the
quotient space ${\bf C}^2/{\bf Z}_2$ everywhere except at the origin,
where there is a copy of the classifying stack $B{\bf Z}_2$ inserted, 
which mathematically
desingularizes the quotient space.  
In that example, one has a ${\bf Z}_2$ gerbe over the origin
(as $B{\bf Z}_2$ is a gerbe over a point), but nowhere
else.  We have already seen such structures in moduli stacks of flat
connections, and they can also arise in moduli stacks of spaces:  for example,
the moduli stack of elliptic curves admits special points which are
locally quotients, and so have finite gerbe structures.  (The elliptic
curves at those points have automorphisms not possessed by generic
elliptic curves.)  Sometimes
(though not always) a gerbe structure at a subvariety on a moduli space
will reflect an orbifold structure.  For example, the moduli space of
K3 surfaces contains a ${\bf Z}_2$ orbifold point, at which the K3 
is represented by $[T^4/{\bf Z}_2]$.  In this example, the 
orbifold structure on the moduli space reflects the quantum symmetry of
the orbifold theory (though as already noted, this is not always the case).

In passing, we should also mention that there may be further examples
of string theories with gerby moduli spaces implicit in \cite{vafa-geom},
which `geometrically engineers' four-dimensional theories with nonabelian
gauge groups from type II compactifications on singular spaces.

\section{Topological defects and gerby moduli spaces}
\label{top-defects}

Recall stable topological defects are classified by the homotopy groups
of the moduli space $M$:  cosmic strings\footnote{
For example, the stringy cosmic strings of \cite{stringycs} arise
from the fact that $\pi_1$ of the moduli stack
of elliptic curves
is $SL(2,{\bf Z})$.  
(This stack should be distinguished from its Deligne-Mumford compactification.
That compactification maps onto $S^2$, hence its homotopy groups have all
of the complexity of the homotopy groups of $S^2$ and more \cite{tonypriv}.)
However, the higher homotopy groups all vanish, 
so from compactifications on elliptic curves, the only topological
defects one can get are stringy cosmic strings.
} arise from $\pi_1(M)$,
monopoles from $\pi_2(M)$, textures from $\pi_3(M)$.
Homotopy groups can be defined for moduli stacks (see {\it e.g.}
\cite{kresch1} and references therein), and in particular
for moduli spaces with gerbe structures, and are not quite the same
as the homotopy groups of the underlying spaces.  In this section we will
outline such homotopy and discuss their potential
application to topological defects.

Let us begin by outlining some pertinent facts about homotopy of
gerbes.
For  ${\cal M}$ a $G$-gerbe\footnote{
In the special case that the gerbe is trivial, {\it i.e.}
${\cal M} =M \times BG = [M/G]$ for
trivially-acting $G$,
$M$ is a $G$-bundle over ${\cal M}$, and so
there is an additional long exact sequence
\begin{displaymath}
\cdots \: \longrightarrow \: \pi_n(G) \: \longrightarrow \: 
\pi_n(M) \: \longrightarrow \: \pi_n({\cal M}) \: \longrightarrow \:
\pi_{n-1}(G) \: \longrightarrow \: \cdots.
\end{displaymath}
} over
$M$, there is a homotopy long exact sequence
\begin{displaymath}
\cdots \: \longrightarrow \: \pi_n(BG) \: \longrightarrow \: \pi_n({\cal M})
\: \longrightarrow \: \pi_n(M) \: \longrightarrow \: \pi_{n-1}(BG) \:
\longrightarrow \: \cdots.
\end{displaymath}
In the sequence above, $BG$ denotes the ``classifying stack'' of $G$
(so named because of formal similarities with the classifying space of $G$).
Technically, the classifying stack is defined as
\begin{displaymath}
BG \: \equiv \: \left[ \frac{\rm point}{G} \right].
\end{displaymath}  
If we think of a gerbe as being analogous to a fiber bundle, then the
fibers are copies of $BG$.  In terms of homotopy groups, it can be shown that
$\pi_n(BG) = \pi_{n-1}(G)$.
In particular,
for finite $G$, $\pi_1(BG) = G$.
(As maps from a space $X \rightarrow BG$ are defined by principal $G$-bundles
over $X$, and principal $G$-bundles over $S^1$ are classified by
conjugacy classes in $G$, not elements of $G$, this also means we should
be careful about interpreting $\pi_1$ as a group of maps from a circle.)

Now, let us apply the description above to cosmic strings,
and discuss whether topological defects should be classified by homotopy
groups of the gerbe or the underlying space.

Consider, for example, a moduli space with a ${\bf Z}_n$ gerbe
structure, call it ${\cal M}$.   
If we denote the underlying (technically, ``coarse'') moduli space by $M$,
then topological defects would ordinarily be computed by homotopy of
$M$.  The effect of the gerbe is to add a $B{\bf Z}_n$ fiber
over each point of $M$.
Over any point of $M$, therefore, is a copy of $B{\bf Z}_n$,
which has $\pi_1 = {\bf Z}_n$.  

If topological defects are classified by homotopy groups of the moduli
stack, not the underlying moduli space $M$, then we would get
a cosmic
string defined by a loop around $B {\bf Z}_n$ fibers, which may or may not
be globally stable depending upon global properties\footnote{
Suppose, for example,
the gerby moduli space ${\cal M}$
is the nontrivial ${\bf Z}_n$ gerbe over ${\bf P}^1$
defined by taking two homogeneous coordinates $x$, $y$ to have weight
$n$ under ${\bf C}^{\times}$.  
The space $S^3$ is a circle bundle over
this gerbe, so we have a homotopy long exact sequence of the form
\begin{displaymath}
\cdots \: \longrightarrow \: \pi_n(S^1) \: \longrightarrow \:
\pi_n(S^3) \: \longrightarrow \: \pi_n({\cal M}) \: \longrightarrow \:
\pi_{n-1}(S^1) \: \longrightarrow \: \cdots.
\end{displaymath}
In particular, since $\pi_0(S^1) \cong \pi_0(S^3)$ and
$\pi_1(S^3) = 0$, we have that $\pi_1({\cal M}) = 0$,
and so the gerbe ${\cal M}$ is simply connected.
Thus, our hypothetical cosmic string would not be globally stable.
} of the gerbe.
This would be some new type of cosmic string, as ordinary cosmic strings
arise from $\pi_1(M)$.  In this new type of cosmic string, the moduli space
scalars would be unchanged as one walks around the string, except that the
theory would undergo some ${\bf Z}_n$ gauge transformation around such
a loop.  Only massive noninvariant fields would see that gauge
transformation.  

Let us now turn to physical examples.  The hypothetical cosmic string
above sounds very similar to the ${\bf Z}_n$ cosmic string discussed
in {\it e.g.} \cite{vs}[section 4.2.2].  There, one has an $SU(2)$ gauge
theory with a pair of triplet-valued Higgs fields which are required
(by virtue of a potential term) to be orthogonal.  Giving the first
Higgs triplet a vev breaks $SU(2)$ to $U(1)$; giving the second
an (orthogonal) vev breaks $U(1)$ to ${\bf Z}_2$.
After both symmetry breakings have occurred, one has ${\bf Z}_2$
cosmic strings, as $\pi_1( SU(2) / {\bf Z}_2 ) = {\bf Z}_2$.
In such a theory, the moduli space of possible Higgs vevs has a natural
${\bf Z}_2$ gerbe structure, and the cosmic strings described by
\cite{vs} seem to naturally coincide with the cosmic strings we have
outlined above arising from homotopy of the gerbe.  In fact, 
our homotopy considerations would appear to give a new perspective
on the ${\bf Z}_n$ cosmic strings of \cite{vs}, as they are discussed
there only as homotopy of group cosets, and here we seem to have
found the same structure in homotopy of a moduli stack.

Unfortunately, further analysis does not seem to bear out this perspective.
One seeming counterexample arises in \cite{sy}[section 4.2].
That reference also describes ${\bf Z}_n$ cosmic strings, though in that
case, the adjoints act primarily as spectators, and the cosmic string
solution naturally involves winding of vevs of massless fundamentals,
with a potential fixing their vevs to be nonzero.
In the present case, for physical relevance of a gerbe structure, we need
noninvariant fields, albeit massive noninvariant fields.  If the 
noninvariant fields are massive, then its vev vanishes, and any
sort of winding solution of the form outlined in 
\cite{sy}[section 4.2], unlikely.

Here is a more convincing counterexample. 
Consider an $SU(2)$ gauge theory containing only a single
Higgs triplet, the $SU(2)$ would only be broken to $U(1)$,
and although the resulting theory has monopoles (as 
$\pi_2( SU(2)/U(1)) = {\bf Z}$), it does not have cosmic strings
(as $\pi_1( SU(2)/U(1)) = 0$).  Thus, in this case, the homotopy of the
gerbe gives a misleading result.

One potential fix to the counterexample above is to replace
Deligne-Mumford stacks with more general Artin stacks (which are not
required to have finite stabilizers).  In the example above,
an $SU(2)$ gauge theory with a single Higgs triplet, an Artin moduli stack
would naturally have a $U(1)$ gerbe structure.  
Now, $\pi_2( BU(1)) = \pi_1( U(1)) = {\bf Z}$, so the same homotopy
analysis of the gerbe would imply the existence of monopoles in this
example, matching the physical result.  For that matter, as
$\pi_1( BU(1) ) = \pi_0(U(1)) = 0$, there is no prediction of
cosmic strings, also matching the physics.
On the other hand, as the gerbe structure would only see the unbroken 
part of the gauge group, not the original gauge group, it seems unlikely
that the example above would generalize to give accurate results in
other cases.

Our tentative
conclusion is that, at least for Deligne-Mumford moduli stacks,
the homotopy of the gerbe is misleading, the
extra elements of $\pi_1$ that one encounters do not reflect physically
meaningful new cosmic string solutions, and that topological defects
should be counted by homotopy of the underlying space.
This then begs the question of how to understand cosmic strings and
other topological defects when the moduli space is a more complicated
stack.

For completeness, let us also formally discuss higher defects
in Deligne-Mumford stacks, though as already established, their
physical relevance may not be significant.
To be specific, consider the ${\bf Z}_n$ gerbe
over ${\bf P}^1$ defined by taking two homogeneous coordinates $x$, $y$
to have weight $n$ under ${\bf C}^{\times}$, rather than weight $1$.
Call this gerbe ${\cal M}$.
We shall consider hypothetical 
monopoles arising from the moduli stack ${\cal M}$. 
From the long exact sequence for homotopy, we have
\begin{displaymath}
\pi_2(B {\bf Z}_n) \: \longrightarrow \: \pi_2({\cal M}) \: \longrightarrow
\: \pi_2({\bf P}^1) \: \longrightarrow \: \pi_1(B{\bf Z}_n) \:
\longrightarrow \: \pi_1({\cal G}).
\end{displaymath}
Now, $\pi_2(B{\bf Z}_n) = 0$, and
it can be shown $\pi_1({\cal M}) = 0$, so we have
\begin{displaymath}
0 \: \longrightarrow \: \pi_2({\cal M}) \: \longrightarrow \:
\pi_2({\bf P}^1) \: \longrightarrow \: {\bf Z}_n \: \longrightarrow \: 0.
\end{displaymath}
Thus, the total number of stringy monopoles
arising from this gerby moduli space would be countable, just as for an
ordinary projective space, but note that not every monopole arising
from ${\bf P}^1$ arises when the moduli stack is a gerbe over ${\bf P}^1$,
closely mirroring the fact that in two-dimensional sigma models on gerbes
there is a restriction on degrees of allowed maps.

We leave for future work a detailed discussion of global topological defects
for more general moduli stacks.  Our results here suggest that global
gerbe structures may not be relevant, at least for
Deligne-Mumford moduli stacks.  It is possible that this is ultimately a 
reflection of subtleties in low-energy effective actions discussed
in section~\ref{4dphys}.  We shall not attempt to address
the relevance of homotopy of gerbe structures that exist only over
subvarieties, or homotopy of Artin moduli stacks.

In passing, we should mention that \cite{banks-seib}[section 4.2] 
speculated on the existence of cosmic strings of the form above in
cases with gerby moduli spaces.

\section{Consistency conditions on classical supergravity}
\label{sugrav-stacks}

In this section we will discuss consistency conditions on 
classical supergravities.  We begin by reviewing results 
\cite{bag-ed1,nati0,git-sugrav} for the case that the moduli space
is a smooth manifold, and then we generalize to smooth Deligne-Mumford
stacks, focusing on gerbes over manifolds.

\subsection{Review of standard supergravity case}

First, let us recall the argument of Bagger and Witten \cite{bag-ed1} that the
K\"ahler class of the moduli space of scalars of
a supergravity theory is quantized, in the case that that moduli space
is a smooth manifold.  First, recall that across coordinate
patches on the moduli space, the K\"ahler potential $K$ transforms as
\begin{displaymath}
K \: \mapsto \: K \: + \: f \: + \: \overline{f},
\end{displaymath}
where $f$ is a holomorphic function of moduli,
which must be accompanied by a rotation of the
gravitino $\psi_{\mu}$
and the superpartners $\chi^i$ of the
scalar fields on the moduli space:
\begin{equation}  \label{grav-trans}
\chi^i \: \mapsto \: \exp\left( + \frac{i}{2} {\rm Im}\, f
\right) \chi^i, \: \: \:
\psi_{\mu}  \: \mapsto \: \exp\left( - \frac{i}{2}{\rm Im}\, f
\right) \psi_{\mu}.
\end{equation}
Consistency of the rotations~(\ref{grav-trans}) across triple overlaps
(even within classical physics) implies that
the $f$'s define a line bundle with even $c_1$.
If we denote that line bundle by ${\cal L}^{\otimes 2}$, 
then the gravitino is a spinor-valued section of $TX \otimes
\phi^* \mathcal{L}^{-1}$, where $X$ is the four-dimensional 
low-energy effective
spacetime and $\phi: X \rightarrow  M$ the boson of the four-dimensional
nonlinear sigma model on the compactification moduli space
$M$, and that the fermions $\chi^i$ are spinor-valued sections of
$\phi^*(T  M \otimes \mathcal{L})$.  Similarly, one shows that the
superpotential is a holomorphic section of ${\mathcal L}^{\otimes 2}$,
and, in order to have a positive-definite metric, 
${\mathcal L}^{-2}$ 
(whose $c_1$ matches the cohomology class of the K\"ahler form) 
must be ample\footnote{
Specifically, positive-definiteness of the metric implies that
every closed analytic subvariety of the moduli space $M$ has positive volume
with respect to $c_1({\mathcal L}^{-2})$, {\it i.e.} for $Y \subset M$ closed
of dimension $p$,
\begin{displaymath}
\int_Y c_1({\mathcal L}^{-2})^p \: > \: 0
\end{displaymath}
which is equivalent to ${\mathcal L}^{-2}$ being ample.
}.

The recent paper \cite{git-sugrav} extended the analysis of
\cite{bag-ed1} to gauged group actions.
If we gauge the action of some group $G$ on the target space of the
nonlinear sigma model in the supergravity theory, then we have to
lift that group action to the Bagger-Witten line bundle
${\cal L}^{\otimes 2}$ in order to define the gauging globally.  
We can see this explicitly in the supergravity gauge transformations.
Under an infinitesimal group action 
\begin{displaymath}
\delta \phi^i \: = \: \epsilon^{(a)} X^{(a) i} 
\end{displaymath}
where $X^{(a)}$ is a holomorphic Killing vector describing the
infinitesimal group action,
the superpartners $\chi^i$, gaugino $\lambda^{(a)}$, and gravitino
$\psi_{\mu}$ transform as
\begin{eqnarray*}
\delta \chi^i & = & \epsilon^{(a)} \left(
\frac{\partial X^{(a) i} }{\partial \phi^j} \chi^j \: + \:
\frac{i}{2} {\rm Im} \: F^{(a)} \chi^i \right), \\
\delta \lambda^{(a)} & =& f^{abc} \epsilon^{(b)} \lambda^{(c)}
\: - \: \frac{i}{2} \epsilon^{(a)} {\rm Im} \: F^{(a)} \lambda^{(a)}, \\
\delta \psi_{\mu} & = & - \frac{i}{2} \epsilon^{(a)} {\rm Im} \: F^{(a)}
\psi_{\mu},
\end{eqnarray*}
where $F^{(a)} = X^{(a)} K + i D^{(a)}$ ($K$ the K\"ahler potential),
and $F^{(a)}$ is easily checked to be
holomorphic.
For real $\epsilon^{(a)}$,
the K\"ahler potential undergoes a standard K\"ahler transformation
\begin{displaymath}
\delta K \: = \: \epsilon^{(a)} F^{(a)} \: + \: 
\epsilon^{(a)} \overline{F}^{(a)},
\end{displaymath}
hence in the gauge transformations above,
terms proportional to ${\rm Im}\: F^{(a)}$ are precisely encoding
the K\"ahler transformations on fermions given in equation~(\ref{grav-trans}).
Thus, the gauge-transformation terms proportional to ${\rm Im} \: F^{(a)}$
are encoding an infinitesimal lift
of the group action to $\mathcal{L}$.

To define the gauge theory, we must extend the infinitesimal action encoded
in supergravity to an action of the group, not just the Lie
algebra.
In general, lifts of group actions to line
bundles need neither exist nor be unique.
The existence issue provides a constraint on possible consistent
supergravities.  (Lack of) uniqueness is encoded in the Fayet-Iliopoulos
parameter, as it is argued in \cite{git-sugrav} that implicit in the
supergravity is the statement that the Fayet-Iliopoulos parameter
determines different lifts of the action of $G$ to ${\cal L}^{\otimes 2}$.
As such lifts are quantized, the Fayet-Iliopoulos parameter is necessarily
quantized, and corresponds to an element of ${\rm Hom}(G,U(1))$ for
$G$ the gauge group.
Just as D-terms are understood in rigid supersymmetry in terms of
symplectic quotients, the paper \cite{git-sugrav} argues that the
structure above in supergravity can be understood in terms of
`geometric invariant theory' quotients (see {\it e.g.} 
\cite{git,newstead,kirwan}),
the algebro-geometric analogue of
symplectic quotients.  In particular, in a geometric invariant theory
quotient, the analogue of the Fayet-Iliopoulos parameter is quantized,
because it is realized as a lift of a group action to a line bundle.

In the rest of this section we shall extend the analysis of
\cite{bag-ed1} and \cite{git-sugrav} to smooth Deligne-Mumford stacks,
focusing on gerbes over manifolds.

\subsection{Generalization to smooth Deligne-Mumford stacks}

The original work of Bagger-Witten \cite{bag-ed1} and
followups \cite{git-sugrav}, reviewed above, 
only considered supergravity theories
in which the moduli space is a smooth manifold.  However, moduli spaces
which are smooth manifolds are vanishingly rare -- more typically,
they have singularities and/or various stack structures, and a generalization
of \cite{bag-ed1,git-sugrav} to such cases would be useful.

Formally, generalizing \cite{bag-ed1} and \cite{git-sugrav}
to moduli `spaces' that are smooth\footnote{
Experts should note that since we are implicitly working over the complex
numbers, `smooth' implies, for example, that there are no nonreduced
scheme structures \cite{tonypriv}.  
}
Deligne-Mumford stacks is very straightforward -- the analysis 
of \cite{bag-ed1}, \cite{git-sugrav} applies
with only minimal modification.  The main caveat is that specifying a
``nonlinear sigma model on a stack'' does not uniquely specify the physics;
we must choose a presentation of the stack, and we could get different
physics according to the choice.  Put another way, there are multiple distinct
physical theories, components of possibly multiple supergravities,
that can be interpreted as a 
nonlinear sigma model on a single fixed
stack.  In four dimensions, we use stacks to
provide a `universal' four-dimensional object for which any given physical
realization corresponds to a presentation.  

Let us outline the analysis in two
different presentations:
\begin{itemize}
\item Deligne-Mumford
stacks have coverings by open sets, so first consider a presentation in which
the atlas is such a collection.  Then, we can work patch-by-patch.
Physically, this means we have a nonlinear sigma model on each open set,
with perhaps a discrete gauge group.
Just as for spaces,
transformations of the K\"ahler
potential across coordinate patches imply that there is a line bundle
${\cal L}$ over the moduli space, to which the gravitino $\psi_{\mu}$
and chiral superpartners $\chi^i$ couple.  In other words, just as for the
case that the moduli stack $M$ is a space, the gravitino is a spinor-valued
section of $TX \otimes
\phi^* \mathcal{L}^{-1}$ ($X$ the four-dimensional spacetime, 
$\phi: X \rightarrow M$), and the superpartners $\chi^i$ are spinor-valued
sections of $\phi^*(T  M \otimes \mathcal{L})$.  The superpotential is
a section of ${\mathcal L}^{\otimes 2}$.
The Fayet-Iliopoulos parameter is a choice of lift of
group action to ${\mathcal L}$, and such choices, possible values of the
Fayet-Iliopoulos parameter are elements of ${\rm Hom}(G, U(1))$ for
$G$ the gauge group. 
\item Now, let us consider another presentation.  Nearly
all (see \cite{msx} for a discussion of exceptions) smooth
Deligne-Mumford stacks
can be presented as global quotients of ordinary smooth manifolds by 
(not necessarily finite) groups,
whose actions need not be effective.
To such presentations we can associated gauged supergravities, to which we
can 
immediately apply \cite{git-sugrav}.  To be specific, suppose the
moduli stack
is presented as $[Y/G]$ for some smooth manifold $Y$ and some group $G$,
corresponding physically to a supergravity theory with moduli space $Y$
and gauged\footnote{
The stack does not specify a classical gauge coupling; again, in four
dimensions, we associate stacks to physics but not physics to stacks.
} $G$ action.
In this case, 
the Bagger-Witten line bundle on the cover $Y$ with a 
$G$-equivariant structure (specified when one defines the gauge theory
\cite{git-sugrav}) is equivalent to a (Bagger-Witten) line bundle
on $[Y/G]$.  Other results follow analogously.
For example, in this presentation, quantization of
the K\"ahler form on the stack $[Y/G]$ follows from both quantization of
the K\"ahler form on $Y$ \cite{bag-ed1} and from
quantization of Fayet-Iliopoulos parameters \cite{git-sugrav}.
In all cases, generalizing Bagger-Witten \cite{bag-ed1} to this presentation
intertwines the analyses and results of \cite{bag-ed1} and
\cite{git-sugrav}.  
(The toy example of \cite{nati0} was realized by a presentation of
this form.)
\end{itemize}
There exist more types of presentations of stacks ({\it e.g.} groupoid
quotients), and so possibly more physical theories; in this paper,
we shall discuss only the presentations above.
As emphasized in section~\ref{4dphys}, even in the IR these presentations
can be physically distinct.

Regarding metric positivity, notions of ampleness and corresponding
constraints on stacks are discussed in \cite{kresch1}; we assume, but have
not carefully checked, that they are pertinent here.

In other words, formally,
the results of \cite{bag-ed1,git-sugrav} carry over more or
less immediately to smooth Deligne-Mumford stacks, at least in
presentations of the form above.
The only significant differences
are as follows:
\begin{itemize}
\item A technical point is that cohomology of stacks more naturally
lives in a different stack, the ``associated inertia stack,'' not the
stack itself.  Thus, the analysis of \cite{bag-ed1} still implies that
the cohomology class of the K\"ahler form on the moduli stack should match
the cohomology class of $c_1$ of the Bagger-Witten line bundle on the stack,
but although the K\"ahler form and Bagger-Witten line bundles themselves
live on the stack, the cohomology lives in the associated inertia stack,
and must be compared there.  This
adds no essential physical constraint.
\item Because the stack is, roughly, a space with (finite) 
automorphisms, coordinate
patches need match only up to (finite) automorphisms.  
Hence, for bundles on stacks, transition functions on triple overlaps need
only close up to finite automorphisms.  This means an honest bundle
on a stack can be a `twisted' or `fractional' bundle on a
space -- objects which are not bundles in the ordinary sense.
(We shall define these below.)
Put another way, there are more bundles on stacks than on underlying spaces,
and many things on spaces that are not quite bundles, become honest
bundles on stacks.  Therefore, we need to carefully examine possible
Bagger-Witten line bundles on stacks for possible physical subtleties.
\end{itemize}

Let us examine the second issue above, in the
special case of smooth Deligne-Mumford stacks that have a (finite)
gerbe structure over a smooth manifold.  A twisted bundle on a space 
(see {\it e.g.} \cite{andreithesis,andrei1,andrei2,cks})
is a bundle in which the transition functions close only up to
a higher cocycle; schematically:
\begin{displaymath}
g_{\alpha \beta} g_{\beta \gamma} g_{\gamma \alpha} \: = \: h_{\alpha
\beta \gamma}
\end{displaymath}
for some Cech cocycle $h_{\alpha \beta \gamma}$, where the $g_{\alpha \beta}$
are transition functions.  Consistency requires that the rank of a 
twisted bundle be related to the order of the cohomology element defined
by $(h_{\alpha \beta \gamma})$; since we are interested in line bundles,
no nontrivial twisted bundles can contribute.
Therefore, we need only consider the possibility that the Bagger-Witten
line bundle might be a fractional line bundle.

To explain fractional line bundles, which will play a crucial role in 
this section,
let us give an explicit example.
Consider\footnote{
This example could not arise physically because of anomalies.  We give it
here as a purely mathematical demonstration and explanation of
fractional line bundles, no more.
} a ${\bf Z}_k$ gerbe on ${\bf P}^n$ defined by
$n+1$ homogeneous coordinates of weight $k$, {\it i.e.}
\begin{displaymath}
\left[ \frac{ {\bf C}^{n+1} - 0 }{ {\bf C}^{\times} } \right],
\end{displaymath}
where the ${\bf C}^{\times}$ acts as
\begin{displaymath}
\left( x_0, \cdots, x_n \right) \: \mapsto \:
\left( \lambda^k x_0, \cdots, \lambda^k x_n \right).
\end{displaymath}
On this gerbe one can define line bundles with arbitrary weight under
the ${\bf C}^{\times}$.  For example, a line bundle of weight $m$ has
total space\footnote{
Curiously, total spaces of fractional line bundles over gerbes often have the
property that they are honest spaces with orbifolds, instead of gerbes,
as is implicit in the expression given.  Despite the existence of
the orbifold structure along the zero section,
one still has a notion of local trivializations; over the gerbe, the
total spaces of the fractional bundles above have a local description
of the form $U \times B {\bf Z}_k \times {\bf C}$ for $U$ an open patch
on ${\bf P}^n$ and $B {\bf Z}_k$ the classifying stack of ${\bf Z}_k$.
The point is that for any vector space $V$, the quotient $[V/G]$ is
the same stack as the total space of a vector bundle of fiber $V$ over
$BG$.  The difference between the two descriptions might be described
as distinguishing fibers over `gerby points' from fibers over `variety points':
in the former case,
one speaks of a vector bundle over $BG$, whereas in the latter, one speaks of
$[V/G]$.  Put another way, the (representable) projection map to
the gerbe on ${\bf P}^n$ has two types of fibers:  the fiber over
a point of ${\bf P}^n$ is
\begin{displaymath}
\left[ \frac{ {\bf C}^{\times}_k \times {\bf C}_m }{ {\bf C}^{\times} } \right]
\end{displaymath}
(with subscripts indicating weights), whereas the fiber over a $B {\bf Z}_k$
is just ${\bf C}$.
}
\begin{displaymath}
\left[ \frac{ ( {\bf C}^{n+1} - 0) \times {\bf C} }{ {\bf C}^{\times} } 
\right],
\end{displaymath}
where the ${\bf C}^{\times}$ acts as
\begin{displaymath}
\left( x_0, \cdots, x_n; y \right) \: \mapsto \:
\left( \lambda^k x_0, \cdots, \lambda^k x_n; \lambda^m y \right).
\end{displaymath}
When $m$ is divisible by $k$, this is the pullback of an honest
line bundle on ${\bf P}^n$, namely ${\cal O}(m/k)$.
More generally, the line bundle above on the gerbe is sometimes
(ambiguously) denoted ${\cal O}(m/k)$ even when $m$ is not divisible by $k$.
In such cases, one has a `fractional' line bundle.
(See \cite{tonyme} for a more complete description of fractional line bundles.)

In passing, the properties of fractional branes at orbifold points are
not unrelated to fractional bundles.  Ultimately the reason for the
relationship is that in an orbifold, there is a gerbe structure appearing
over orbifold points, which has the effect of desingularizing the
orbifold.

It should now be clear that these
fractional line bundles on the gerbe are precisely what is
being described in the example of \cite{nati0}, and more
generally whenever one has ``fractional Fayet-Iliopoulos parameters.''
If one were to pick a different physical presentation of the gerby
moduli `space,' say as a nonlinear sigma model with a restriction
on nonperturbative sectors rather than as a gauged linear sigma model,
then the line bundle $L$ of which the superpotential is a section
would just be taken to be a fractional line bundle from the outset.

Now that we have explained fractional line bundles,
let us return to our discussion of them as possible Bagger-Witten line
bundles arising when the moduli stack possesses a (finite) gerbe
structure, and discuss possible consistency conditions.
We see two possibilities:
\begin{enumerate}
\item One possibility is that allowed K\"ahler forms can have
cohomology classes matching the 
(image of the) first Chern class of any line bundle on the gerbe,
including any fractional line bundle.
\item Another possibility is that allowed K\"ahler forms can have
cohomology classes matching the
(image of the) first Chern class only of line bundles which are
pullbacks of line bundles on the underlying space -- no fractional line
bundles allowed.
\end{enumerate}
The recent paper \cite{nati0} argued the former case,
that if the moduli `space' of the
supergravity theory were actually a gerbe over an ordinary space
then the quantization condition of
Bagger-Witten should be modified, and fractional values of the 
Fayet-Iliopoulos parameter should be allowed.
We shall now study this claim in detail.

Let us re-examine the example\footnote{
The supersymmetric ${\bf C}{\bf P}^n$ model would be marginally simpler
to describe, but is also anomalous.
} of a $U(1)$ gauge theory coupled to
supergravity discussed in section~\ref{fourdim-basics},
in the spirit of \cite{nati0}.
In this theory, the supergravity moduli space is ${\bf C}^{2n+2}$,
and under the gauged $U(1)$, the fields have the following charges:
$n$ fields $\phi_i$ of charge $k$, $n$ fields $\tilde{\phi}_i$ of charge $-k$,
one field $\chi$ of charge $+1$, one field $\tilde{\chi}$ of charge $-1$.
Furthermore, the fields of charge $\pm 1$ have mass $m$.

Restricting to the massless fields, the D-term condition has the form
\begin{displaymath}
\sum_i k | \phi_i |^2 \: - \: \sum_i k | \tilde{\phi}_i |^2 \: = \: r
\end{displaymath}
where $r$ is the Fayet-Iliopoulos parameter.  As discussed elsewhere
\cite{git-sugrav,nati0}, in supergravity $r$ is constrained to be an
integer, so \cite{nati0} observed that
when we divide by the common factor of $k$,
the D-term condition becomes
\begin{displaymath}
\sum_i  | \phi_i |^2 \: - \: \sum_i  | \tilde{\phi}_i |^2 \: = \: r/k
\end{displaymath}
formally giving a fractional Fayet-Iliopoulos parameter
(albeit normalized in such a way as
to make that fact less explicit).

Naively,
the model above appears to describe a loophole in the
analysis of Bagger-Witten \cite{bag-ed1}, by allowing for fractionally
quantized metrics.  However, as discussed in section~\ref{4dphys},
the infrared limit of a four-dimensional gauged sigma model (in effective
field theory) need not
be the same as a four-dimensional nonlinear sigma model.
Thus, the example above is not describing a loophole in Bagger-Witten,
as it does not RG flow to a theory of the form analyzed by Bagger-Witten.  
Rather, it is giving
a quantization condition on a different theory than is considered
by Bagger-Witten.  (We would like to thank J.~Distler for emphasizing
this point to us.)

The physics above maps to a stack, 
a ${\bf Z}_k$ gerbe over a space ${\bf C}^{2n}//{\bf C}^{\times}$,
and a fractional line bundle that extracts the `universal' aspects
of the physics above.  The fractional line bundle over the gerbe
corresponds to the equivariant structure implicit in the
choice of Fayet-Iliopoulos parameter.  We cannot consistently construct
a low-energy effective field theory by integrating out the Higgsed
gauge field, and so there is no regime in which we can consistently
talk about a fractionally quantized metric; however, 
we can nevertheless apply stacks to give a `universal' object
encoding some essential aspects of the physics, and the gauge theory
in question would be described mathematically by a fractional line bundle
on a gerbe. 

So far we have discussed the interpretation of 
certain choices of equivariant structures on Bagger-Witten line bundles.
It remains to understand whether those choices that lead to
fractional line bundles on gerbes are physically consistent.

In particular,
let us examine the kinetic terms for the gravitino $\psi_{\mu}$ and
superpartners $\chi^I$ more systematically.  
Recall the gravitino $\psi_{\mu}$ is a spinor-valued section of
$TX \otimes \phi^* {\cal L}^{-1}$, and the fermions $\chi^I$ are
spinor-valued sections of $\phi^* ( T {\cal M} \otimes {\cal L})$,
where $X$ is the four-dimensional low-energy effective spacetime,
and $\phi: X \rightarrow {\cal M}$ is the bosonic map in the four-dimensional
nonlinear sigma model in the supergravity.
If the moduli space admits a gerbe structure, and the Bagger-Witten line bundle
${\cal L}$ is fractional, 
then there are some potential
issues:
\begin{itemize}
\item First, fractional line bundles have no smooth (or even continuous)
single-valued sections.
\item Second, as noted earlier, seen as bundles over the underlying space,
fractional bundles have orbifold singularities in their fibers, making
a metric on those fibers potentially singular.  As that metric appears
in the fermion kinetic terms, this is potentially a hazard.
\end{itemize}
In principle, both of these problems are solved by the fact
that if the moduli space ${\cal M}$ has a gerbe structure,
then the path integral only sums over maps $\phi$ with degrees 
satisfying certain
divisibility properties -- this is one of the defining properties of
a nonlinear sigma model on a gerbe.  (That said, if the four-dimensional
spacetime $X$ is ${\bf R}^4$, then as discussed before this matter is somewhat
trivial, but let us describe the most general case here.)
That same divisibility
criterion ensures that $\phi^* {\cal L}$ is an honest line bundle, not
anything fractional.  As a result, even when ${\cal L}$ is fractional,
the gravitino $\psi_{\mu}$ and fermions $\chi^I$ do exist as single-valued
objects, coupling to honest bundles, with smooth fiber metrics.

In passing, let us mention another potential issue.
If the SCFT moduli space admits a gerbe structure, and the K\"ahler
form arises from a fractional line bundle, then there is an interesting
structure on the worldsheet operators over SCFT moduli space
(see \cite{dist-trieste} for a discussion for ordinary moduli spaces).
Specifically, as we walk around the SCFT moduli space, some of the worldsheet
operators (including the spectral flow operator)
acquire phases from the (fractional) line bundle, and hence are
necessarily multi-valued over the SCFT moduli space.
This is at least odd, though not necessarily a physical contradiction.
For example, the $SL(2,{\bf Z})$ transforms in monodromies on the $u$-plane
in Seiberg-Witten theory tell us that the low-energy effective action there
is really only globally-defined on an $SL(2,{\bf Z})$-Riemann surface
covering the $u$-plane, not the $u$-plane itself.  The situation here
is closely analogous.

As indirect consistency checks that these theories with fractional
Bagger-Witten line bundles are consistent, let us 
point out some closely related (and consistent) examples:
\begin{itemize}
\item One example occurs in two-dimensional (0,2) SCFT's, describing
heterotic strings on gerbes.  If one compactifies a heterotic string on 
a gerbe with a fractional or twisted gauge bundle ({\it i.e.} a bundle
on the gerbe that is not a pullback from the underlying space), the result
looks like a sigma model on a space with a non-honest bundle, and a restriction
on degrees of maps such that the pullbacks of non-honest bundles are
honest.  These will be discussed in detail in
\cite{tonyme}.  (Note the left-moving worldsheet fermions in this example are
closely analogous to the four-dimensional gravitino and so forth we have
been discussing -- both couple to pullbacks of fractional bundles.)  
One way to construct examples is through asymmetric orbifolds,
that act ineffectively on right-movers but effectively on left-movers.
Examples can also be constructed in (0,2) GLSM's, such as 
the (anomaly-free, fractional)
bundle
\begin{displaymath}
0 \: \longrightarrow \: {\cal E} \: \longrightarrow \:
{\cal O}(1)^{\oplus 9} \: \longrightarrow \: 
{\cal O}(9) \: \longrightarrow \: 0
\end{displaymath}
over ${\bf P}^3_{[2,2,2,4]}[10]$, a ${\bf Z}_2$ gerbe over
${\bf P}^3_{[1,1,1,2]}[5]$.
Other two-dimensional examples have been constructed by dimensional
reduction of twisted four-dimensional ${\cal N}=2$ theories, as in
\cite{kap-02}.  These examples all seem to be consistent.
\item It is perhaps worth observing that nonlinear sigma models on total
spaces of fractional bundles are well-behaved.
Consider a (2,2) supersymmetric
gauged linear sigma
model describing 
\begin{displaymath}
\left[ \frac{
( {\bf C}^2 - 0 ) \times {\bf C}
}{ {\bf C}^{\times} } \right]
\end{displaymath}
where the ${\bf C}^{\times}$ acts on $({\bf C}^2 - 0)$ with weight 2,
and on ${\bf C}$ with weight 1, say.  This is the total space of the
(fractional)
line bundle ${\cal O}(1)$ over a ${\bf Z}_2$ gerbe on ${\bf P}^1$;
it is also a modification of the exceptional set away from
the weighted projective stack ${\bf P}^2_{[1, 2, 2]}$.
This is a consistent (2,2) supersymmetric theory.
\item A four-dimensional gauge theory can also be constructed with
closely analogous properties.  Consider an $SU(n)$ gauge theory with
matter in the fundamental of $SU(n)$.  We can interpret this as 
the Feynman diagrams of
$SU(n)/{\bf Z}_n$ gauge theory with a subset of the $SU(n)/{\bf Z}_n$
instantons (omitting fractional instantons), 
restricted so as to make the fundamental matter always
well-defined.  (We cannot precisely call this an $SU(n)/{\bf Z}_n$
gauge theory with a restriction on instantons, because the
$SU(n)/{\bf Z}_n$ gauge transformations are not well-defined on
the matter fields.  For this reason, both the 
$SU(n)$ and the $SU(n)/{\bf Z}_n$ gauge theories obey cluster
decomposition.)  After all, perturbatively an $SU(n)$ and $SU(n)/{\bf Z}_n$
gauge theory are identical (same Lie algebra, same Lagrangian, 
same
Feynman diagrams), the difference between the data given is that an
$SU(n)/{\bf Z}_n$ has additional (`fractional') instantons not present in
the $SU(n)/{\bf Z}_n$ theory.  One could imagine splitting an $SU(n)$
instanton into $SU(n)/{\bf Z}_n$ instantons, but if one does so,
one would have to introduce topological defects in order to allow the
matter in the fundamental representation to be well-defined globally. 
\end{itemize}
This last example perhaps best exemplifies the slogan
\begin{quotation}
Perturbative physics is determined by the Lie {\it algebra} of the gauge
group, but nonperturbative physics is determined by the Lie {\it group},
not just the algebra.
\end{quotation}
mentioned in section~\ref{4dphys}.

Our tentative conclusion is that ``fractional Fayet-Iliopoulos parameters''
are consistent in supergravity theories in which the moduli stack is
a gerbe, and are a reflection of stacky subtleties arising
in more general supergravity theories.  
One should be careful about asserting that this implies a loophole
in Bagger-Witten's old quantization result \cite{bag-ed1},
as the infrared limit of a four-dimensional gauged sigma model 
need not be the same as
a nonlinear sigma model. 
We leave a more detailed
analysis of consistency conditions in supergravity theories with
moduli stacks to future work.

\section{Conclusions}

In this paper we have reviewed recent discussions of quantization of
the Fayet-Iliopoulos parameter in supergravity theories.
We began this paper by reviewing previous work on two-dimensional theories
with restrictions on nonperturbative sectors -- equivalently,
sigma models on gerbes -- and more general aspects of two-dimensional
sigma models on stacks, followed by a discussion subtleties appearing in
four-dimensional analogues.  We gave examples in both field and
string theory of models with gerbe structures on their moduli spaces,
and discussed the action of duality.  We discussed global topological
defects when the moduli space is a stack, focusing on stacks that are
gerbes over smooth manifolds.  We then discussed consistency conditions
on classical supergravity theories for moduli spaces that are smooth
Deligne-Mumford stacks, after reviewing the state-of-the-art for
smooth manifolds.

In the text we listed a number of interesting possible followups.
Another direction that would be interesting to pursue is
sigma model anomalies, in the sense of Moore-Nelson \cite{mn1,mn2,mmn},
in cases where the target space is a gerbe or other stack.

Yet another direction concerns deformation issues.
Briefly, stacks and underlying spaces do not always admit
the same deformations.
To illustrate the principle, consider
a local quotient stack structure resolving an orbifold
singularity on a Calabi-Yau.  (Moduli spaces are typically not
Calabi-Yau, but this will provide a simple example of the deformation
theory issue.)
Although quotient {\it spaces} often admit Calabi-Yau blowups,
corresponding quotient {\it stacks} do not.
(In string compactifications on stacks, this leads to an apparent
mismatch in moduli which was discussed in \cite{msx}.)
Notions of blowup still exist, but are usually not Calabi-Yau.
For moduli stacks appearing in field theory and string theory, then,
a natural question to ask is whether the existence of a quotient
stack structure `resolving' an orbifold singularity on the moduli space
reflects any obstruction to resolution or deformation of the singularity.
It would be interesting to understand if this deformation-theoretic
mismatch had any applications in either field or string theory.

There are several other 
potential applications of such gerbe and stack structures
in field theory moduli spaces that we can imagine.
For example, it would be interesting to understand whether
`stacky' resolutions of quotient singularities on moduli spaces,
{\it i.e.} $[ {\bf C}^n/G ]$ versus ${\bf C}^n/G$,
convey any additional information about the theory, such as properties
of light particles.
It would also be interesting if gerbe structures could be used to
help disentangle confusing potential Seiberg duals.  Examples of
such are discussed in, for example,
\cite{bci}, and there is a gerbe structure on some of the moduli spaces
of the field theories discussed there.
Similarly, it would be interesting to understand the three-dimensional
`mirrors' \cite{is} to theories with nonminimally-charged electrons.
In two dimensions, such mirrors turned out to involve either
discrete-valued fields \cite{glsm} or, equivalently, disconnected targets
\cite{hhpsa}.

It would also be interesting to understand if the ideas in this paper could
be applied to understand the distinctions between $SU(2)$ and $SO(3)$
Donaldson and related mathematical invariants, see {\it e.g.}
\cite{malmen} and references therein.

\section{Acknowledgements}

Some of the ideas in this paper have germinated over the last decade
and been discussed with numerous people, more than we can list here.
In particular,
we would like to thank J.~Distler for numerous 
discussions of cluster decomposition
and supergravity moduli space issues,
A.~Knutson for many discussions of
group theory pertinent to heterotic strings,
T.~Pantev for many years of discussions and
collaboration on strings propagating on gerbes and stacks, and
M.~Strassler for insights and references
on four-dimensional gauge theoretic analogues provided during a visit
to the University of Washington in spring 2005.
We would also like to thank P.~Argyres, A.~Shapere, and M.~Unsal for
more recent useful conversations.

The work of S.H. was supported by the WorldPremier International Research      
Center Initiative, MEXT, Japan, and by a Grant-in-Aid 
for Scientific Research (22740153) from the Japan Society for Promotion      
of Science                                                                      
(JSPS).
E.S. was partially supported by NSF grants
DMS-0705381 and PHY-0755614.

\appendix

\section{Four-dimensional decomposition conjecture}
\label{4d-decomp}

In this appendix we will discuss a four-dimensional analogue of the
decomposition conjecture for two-dimensional CFT's discussed in
\cite{hhpsa}.  This will arise via restricting four-dimensional instantons
(mathematically, $c_2$'s, not $c_1$'s), and so will not be directly
relevant for the gerbes studied elsewhere in this paper.

Consider a four-dimensional SCFT obtained from a gauge theory,
{\it e.g.} ${\cal N}=4$ $SU(n)$ SYM, or ${\cal N}=2$ $SU(n)$ SYM with
$2n$ hypermultiplets in the fundamental, or one of the 
${\cal N}=1$ SCFT's.

In that gauge theory, restrict the nonperturbative sector to instantons
of degree divisible by $k$.  Note that the resulting theory will
not be associated to gerbes -- we are here imposing a restriction on
Pontryagin classes of bundles, visible to four-dimensional
theta angles, whereas gerbe structures would only affect analogues of the
first Chern class.  This theory automatically violates
cluster decomposition, by Weinberg's ancient argument; 
we shall describe how it can be written formally
as a sum of other theories with rotating theta angle.

In this theory, since the instantons have degrees divisible by $k$,
the Chern-Simons vacua split into $k$ separate sets.  The allowed instantons
define tunnelling only between Chern-Simons vacua within the same set.
In this fashion, one recovers $k$ separate zero-energy ground states.
Under the assumption that when the gauge field is extended flatly in
extra dimensions, the Chern-Simons number is cobordism invariant,
the Chern-Simons number is conserved modulo $k$.

Using the state-operator correspondence for conformally-invariant theories,
one can build $k$ different zero-energy states, which for the reasons above
obey the same multiplicative rules as twist fields in the two-dimensional
theories discussed in \cite{hhpsa}, and hence can be used to define
projection operators.

Thus, we conjecture that the four-dimensional SCFT above with theta
angle $\theta$, can be decomposed into a sum of $k$ SCFT's,
and we further conjecture that those $k$ SCFT's are copies of the
SCFT with theta angles $\theta + 2 \pi n/k$ for $n=0, \cdots, k-1$,
where $\theta$ has period $2 \pi$ in the theory where instanton number 1
configurations are allowed.
(This sum has the effect of cancelling out gauge field configurations in
the path integral whose instanton degrees are not multiples of $k$.)

For the two-dimensional decomposition conjecture pertinent to sigma models
on gerbes, there is now abundant evidence, including all-genera partition
function computations in orbifold examples
\cite{hhpsa}, checks in mirror symmetry and quantum cohomology 
\cite{hhpsa}, applications to gauged linear sigma models
\cite{cdhps}, and now checks of predictions for
Gromov-Witten invariants \cite{ajt1,ajt2,ajt3,t1,gt1,xt1}.
By contrast, in the four-dimensional case above, we have no independent 
evidence, no examples,
only the arguments above.

\section{Two-dimensional $BF$ theory and cluster decomposition}
\label{2dbf}

In this section we will examine $BF$ theory in two dimensions,
as an example 
of a manifestly local theory that does not obey cluster decomposition.
Let $B$ be a circle-valued scalar, {\it i.e.} identified under $B \mapsto B + 2 \pi$.
Let $A$ be an abelian gauge field with the usual gauge transformation,
so that locally
\begin{eqnarray*}
A & \mapsto & A \: + \: d \chi, \\
F & \equiv & dA,
\end{eqnarray*}
where $\chi$ is a circle-valued gauge parameter:  $\chi \equiv \chi + 2 \pi$.
Then the field strength $F$ then satisfies the Dirac quantization condition,
\begin{displaymath}
\int F \: \in \: 2 \pi {\bf Z}.
\end{displaymath}

The action for $BF$ theory is
\begin{displaymath}
S \: = \: \frac{k}{2 \pi } \int B F
\end{displaymath}
and the Euclidean action is
\begin{displaymath}
S\ll{\rm E} \: = \: \frac{i\cc k}{2 \pi } \int B F.
\end{displaymath}
This theory is simple
enough that it can be solved exactly and explicitly.  To do this we solve for
the dimension of the Hilbert space of
states on a spatial slice $S\uu 1$ and for the action of the
operator algebra on that Hilbert space.  The particular point
to which we draw attention is the
absence of cluster
decomposition: inside the local operator algebra of the theory is
a pair of local operators
$\co\ll{\pm 1}$ that disobey the condition for cluster decomposition,
in the sense that
\begin{displaymath}
\lim_{x \rightarrow \infty} \langle {\cal O}_1(x) {\cal O}_{-1}(0) \rangle
\: \neq \:
\lim_{x \rightarrow \infty} \langle {\cal O}_1(x) \rangle
\langle {\cal O}_{-1}(0) \rangle.
\end{displaymath}

\heading{Hilbert space of the BF theory}

First, we compute the
overall dimension of the Hilbert space of states on $S\uu 1$.  To do this, we
compute the
partition function on a spatial circle at finite temperature $\b\uu{-1}$:
\bbb
Z(\b) = \sum\ll k \cc \exp{\cc \{- \b E\ll k \}}\ .
\eee
Since the two-dimensional
metric does not appear in the $BF$ action, we expect that the theory is
topological and that the
energies $E\ll k$ should all vanish identically, and that the partition
function $Z(\b)$ is therefore
independent of $\b$.  We will see that this is indeed the case.

To compute the partition
function at finite temperature $\b\uu{-1}$, we perform the path integral
in Euclidean signature with Euclidean time compactified with a radius of $r\ll 2
\equiv {{\b}\over{2\pi}}$.  We also
compactify the
spatial direction with radius $r\ll 1$, so that the Hilbert space becomes
manifestly separable.  We
have then reduced the finite-temperature partition function to a
path integral over a
discretely infinite set of variables, the Fourier modes of the $B$ field
and the $U(1)$ gauge connection.
The path integral over the
nonzero modes is purely Gaussian, and can be performed straightforwardly
so long as we divide appropriately
by the measure for the local $U(1)$ gauge group.  The path integral
over the zero modes we perform separately.

\heading{Path integral measure on a finite torus}

Path integrals in finite volume require a
bit of care in order
to get the overall normalization correct -- we mostly follow the method of
\cite{pathintegral}, deviating from the presentation there
only in details particular to the application here.

Define the measure -- for the gauge group, the gauge field, and the
$B$ field -- as in \cite{pathintegral},
in a local way.  To do this, decompose the fields and the
gauge parameter into normal modes:
\begin{eqnarray*}
A_i(x)  & \equiv & \sum_M a^{(M)} \phi\ll{i,(M)}(x) \\
B(x) & \equiv & \sum_N b^{(N)} \phi_{(N)}(x) \\
\chi(x) & \equiv & \sum_N c^{(N)} \phi_{(N)}(x)
\end{eqnarray*}
where $\phi_{(N)}$ is a set of unit-orthonormalized real eigenfunctions of the
scalar Laplacian, and $\phi_{i,(N)}$
is a set of unit-orthonormalized real vector eigenfields of the
vector Laplacian:
\bbb
\int \cc d\sqd x\cc \sqrt{g} \cc \phi_{(N)}(x)\phi_{(N\pr)}(x)  = \d\ll{N\pr N} ,
\llsk\llsk
\int \cc d\sqd x\cc \sqrt{g} \cc g\uu{ij}\cc
\phi_{i,(M)}(x)\cc \phi_{j,(M\pr)}(x) = \d\ll{M\pr M} \ ,
\eee
and the $a\ll{(M)}, b\ll{(N)}, c\ll{(N)}$
are mode amplitudes, the Fourier transforms of the
dynamical fields.  Then define the path integral measures
\begin{eqnarray*}
DA & \equiv & \prod_{(M)} d a^{(M)} \\
DB & \equiv & \prod_{(N)} d b^{(N)} \\
D\chi & \equiv & \prod_{(N)} d c^{(N)}
\end{eqnarray*}
with unit normalization.

Concretely, for the torus,
we can let $N$ run over the values $0$ and $([p], {\rm re})$,
where $([p],{\rm im})$.  Here, the symbol $[p]$ represents a
pair $\{p\ll i,-p\ll i \}$ of equal and opposite nonzero momenta
obeying appropriate quantization conditions, and ${\rm re}$ and
${\rm im}$ represent the real and imaginary parts of the
mometum eigenfunction.

As for $M$, we let it run over nonzero modes
$([p], {\rm re}, \perp)$, $([p], {\rm im}, \perp)$,
$([p], {\rm re},\parallel)$ and $([p], {\rm im},\parallel)$,
and also over zero modes labelled $(0,I)$.
Here $[p]$ means the same as it does for the scalar eigenmodes,
\it i.e., \rm a pair of equal and opposite nonzero momenta on the two-torus,
and ${\rm re}$ and ${\rm im}$ represent the real and imaginary parts of
a plane wave.  The symbols $\perp$ and $\parallel$ denote the transverse and
longitudinal polarizations for the nonzero modes, and the
$I$ labelling the zero modes
(Wilson lines) runs over the two directions of the torus.

\heading{Path integral over nonzero modes}

Let us now perform the path integral over nonzero modes.  For
a given pair $[p]$, we have two multiplicative contributions
to $Z(r\ll 1, r\ll 2)$: First, we have the
Gaussian path integral over the nonzero
modes $b\ll{([p], {\rm re})}, b\ll{([p], {\rm im})}$
and $a\ll{([p], {\rm re}, \perp)},
 a\ll{([p], {\rm im}, \perp)}$.  And then, we also have
  the Jacobian determinant of the gauge transformation of
 $a\ll{([p], {\rm re}, \parallel)},
 a\ll{([p], {\rm im}, \parallel)}$
by $c\ll{([p], {\rm re})}, c\ll{([p], {\rm im})}$.  First, we
combine the real and imaginary
parts of the plane waves into the natural complex combinations:
 \bbb
 {\Phi}\ll {i,([p], \perp)} & \equiv &
\cc {1\over{\sqrt{2}}} \cc ( \phi\ll{i,([p], {\rm re}, \perp)} + i\cc
 \phi\ll{i,([p], {\rm im}, \perp)}),
 \\
 {\Phi}\ll {i,([p], \parallel)} & \equiv &
\cc {1\over{\sqrt{2}}} \cc ( \phi\ll{i,([p], {\rm re}, \parallel)} + i\cc
 \phi\ll{i,([p], {\rm im}, \parallel)}),
 \\
 {\Phi}\ll {[p]} & \equiv &
\cc {1\over{\sqrt{2}}} \cc ( \phi\ll{([p], {\rm re})} + i\cc
 \phi\ll{([p], {\rm im})}),
 \\
 {\bf a}\ll {([p], \perp)} & \equiv &
\cc {1\over{\sqrt{2}}} \cc ( a\ll{([p], {\rm re}, \perp)} + i\cc
 a\ll{([p], {\rm im}, \perp)}),
 \\
 {\bf a}\ll {([p], \parallel)} & \equiv &
\cc {1\over{\sqrt{2}}} \cc ( a\ll{([p], {\rm re}, \parallel)} + i\cc
 a\ll{([p], {\rm im}, \parallel)}),
 \\
 {\bf b}\ll {[p]} & \equiv &
\cc {1\over{\sqrt{2}}} \cc ( b\ll{([p], {\rm re})} + i\cc
 b\ll{([p], {\rm im})}),
 \\
 {\bf c}\ll {[p]} & \equiv &
\cc {1\over{\sqrt{2}}} \cc ( c\ll{([p], {\rm re})} + i\cc
 c\ll{([p], {\rm im})}).
 \eee
In terms of the complex combinations above, the measure is
\bbb
d a\ll{{\rm re}} \cc d a\ll{{\rm im}} & = & 2\cc d\sqd {\bf a} = 2\cc
 d({\rm Re}\cc {\bf a}) \wedge d({\rm Im}\cc {\bf a}),
\nn\\
d b\ll{{\rm re}} \cc d b\ll{{\rm im}} & = &
2\cc d\sqd {\bf b} = 2\cc d({\rm Re}\cc {\bf b})  \wedge
 d({\rm Im}\cc {\bf b}),
\nn\\
d c\ll{{\rm re}} \cc d b\ll{{\rm im}} & = &
2\cc d\sqd {\bf c} = 2\cc d({\rm Re}\cc {\bf c}) \wedge
 d({\rm Im}\cc {\bf c}),
\eee
where we have suppressed
the indices $[p], \perp$ and $\parallel$, and we have used
the standard convention
\bbb
d\sqd {\bf z} \equiv {i\over{2}} d {\bf z}
\wedge d\bar{{\bf z}} = d({\rm Re}{\bf z}) \wedge
d({\rm Im}{\bf z})
\eee
for the measure on a complex variable ${\bf z}$.

The expansion of $A\ll i(x), B(x)$ and $\chi(x)$ in eigenmodes takes the form
\bbb
A\ll i (x)
& = & ({\rm zero~modes}) \\
& & \hspace*{0.25in} + \sum\ll{[p]}
\cc {\bf a}\ll {([p], \perp)}\cc {\Phi}\ll {i,([p], \perp)}
+ {\bf a}\ll {([p], \parallel)}\cc {\Phi}\ll {i,([p], \parallel)}
+ {\bf a}\ll {([p], \perp)}^*\cc  {\Phi}\ll {i,([p], \perp)}^*
+ {\bf a}\ll {([p], \parallel)}^*\cc {\Phi}\ll {i,([p], \parallel)}^*,
\nn\\
B (x) & = & ({\rm zero~mode}) + \sum\ll{[p]}
\cc {\bf b}\ll {[p]}\cc {\Phi}\ll {[p]}
+ \cc {\bf b}\ll {[p]}^*\cc {\Phi}\ll {[p]}^*,
\nn\\
\chi(x) & = &
({\rm zero~mode}) + \sum\ll{[p]} \cc {\bf c}\ll {[p]}\cc {\Phi}\ll {[p]}
+ \cc {\bf c}\ll {[p]}^*\cc {\Phi}\ll {[p]}^*.
\nn\eee
The orthonormality conditions for the complex normal modes are
\bbb
\int \cc d\sqd x\cc \sqrt{g} \cc \Phi\ll {[p]}^* \Phi\ll {[p\pr]}  =
\int \cc d\sqd x\cc \sqrt{g}
\cc g\uu{ij} \cc \Phi\ll{i,([p],\perp)}^* \Phi\ll{j,([p\pr],\perp)} & = &
\int \cc d\sqd x\cc \sqrt{g} \cc
g\uu{ij} \cc \Phi\ll{i,([p],\parallel)}^* \Phi\ll{j,([p\pr],\parallel)}
\\
& = & \cc \d\ll{[p],[p\pr]},
\nn\\
 \int \cc d\sqd x\cc \sqrt{g} \cc g\uu{ij} \cc \lrdd
\Phi\ll{i,([p],\perp)}^* \Phi\ll{j,([p\pr],\parallel)}\right .   +  \left .
 \Phi\ll{i,([p\pr],\parallel)}^* \Phi\ll{j,([p],\perp)} \rrdd
 &  = &  \int \cc d\sqd x\cc \sqrt{g}
\cc g\uu{ij} \cc \Phi\ll{i,([p],\perp)} \Phi\ll{j,([p\pr],\parallel)}   =   0,
 \nn\\
 \int \cc d\sqd x\cc \sqrt{g} \cc \Phi\ll {[p]} \Phi\ll {[p\pr]} &  = &
\int \cc d\sqd x\cc \sqrt{g} \cc g\uu{ij}
\cc \Phi\ll{i,([p],\perp)} \Phi\ll{j,([p\pr],\parallel)}  =  0.
\nn\eee
and the transversality conditions are
\bbb
g\uu{ij} \pp\ll i \Phi\ll{j,([p],\perp)}
= \e\uu{ij} \pp\ll i \Phi\ll{j,([p],\parallel)} = 0\ .
\eee
A natural choice for the normal modes is
\bbb
\Phi\ll{[p]} & = &
\sqrt{{1\over{\rm vol\ll 2}}} \cc \exp{ \left \{ i p\cdot x \right \} }\ ,
\nn\\
\Phi\ll{i,([p],\perp)} & = &
\sqrt{{1\over{\rm vol\ll 2}}} \cc \sqrt{{|g|}\over{g\uu{mn} p\ll m p\ll n}}
\cc \e\ll{ik} g\uu{kl} p\ll l \cc \exp{ \left \{ i p\cdot x \right \} } \ ,
\nn\\
\Phi\ll{i,([p],\parallel)}
& = &
\sqrt{{1\over{\rm vol\ll 2}}} \cc \sqrt{{1}\over{g\uu{mn} p\ll m p\ll n}}\cc
p\ll i  \cc \exp{ \left \{ i p\cdot x \right \} } \ .
\eee

Now we want to write the
Euclidean action in terms of the complex normal modes.  The action is
of course a topological invariant, but since the normalization
conditions of the normal modes are written in terms of the metric,
it is useful to write the action as
\bbb
S\ll{\rm E} =
 + {{i k}\over{2\pi}}\cc
\int \cc d\sqd x \cc
\sqrt{|g|}\cc {{\e\uu{ij}}\over{2\sqrt{|g|}}}\cc  B\cc F\ll{ij}\ ,
\eee
In terms of the complex normal modes the action is
\bbb
S\ll{\rm E} & = &  \sum\ll{[p]} S\ll{[p]}
\nn\\
S\ll{[p]} & \equiv &
 + {{ k}\over{2\pi}}\cc
|p|\cc \lrdd   {\bf b}\ll{[p]}   {\bf a}
\ll{([p], \perp)}^* \cc  -  {\bf b}\ll{[p]}\uu *  {\bf a}
\ll{([p], \perp)} \rrdd
\nn\\
& = & ({\bf b}\ll{[p]}\uu *   \cc  {\bf a} \ll{([p], \perp)}^*) \cc \co \cc
\lrdd \begin{matrix}{ {\bf b}\ll{[p]}
\cr {\bf a} \ll{([p], \perp )} } \end{matrix} \rrdd\ ,
\nn\eee
where $\co$ is the anti-Hermitean matrix
\bbb
\co\equiv \cc
  - {{i k |p|}\over{2\pi}}
  \cdot \s\uu 2\ , \llsk\llsk
|p|\equiv \sqrt{g\uu{ij} p\ll i p\ll j}\ .
\eee
Since the operator $\co$ has
imaginary eigenvalues, we can define the path integral by the prescription
\bbb
\co\to \co + \e\cc |p|\sqd\cdot {\bf 1}\ll{2\times 2}\ ,
\eee
and let $\e\to 0\uu +$ be a positive real parameter approaching zero from above.
With this definition the action is still local and reduces
in the $\e\to 0\uu +$ limit to the undeformed action,
but the path integral over each set of modes is convergent.

For an operator $\co$ with positive definite real part, the Gaussian
path integral over a vector of complex variables ${\cal A}$ is given by
\bbb
\int \cc \prod\ll{(N)} ( 2 \cc d\sqd  {\cal A}\ll{(N)})
\cc \exp{\left \{ - {\cal A}\dag \cc \co \cc {\cal A} \right \} }
= \lsqq {\rm det}\lrdd {{\co}\over{2\pi}} \rrdd \cc \rsqq\uu{-1}\ ,
\eee
so, letting
\bbb
{\cal A} \equiv \lrdd \begin{matrix}{ {\bf b}\ll{[p]}
\cr {\bf a} \ll{([p], \perp)} } \end{matrix}
\rrdd\ ,
\nn\\
{\cal O}\equiv - {{i k |p|}\over{2\pi}} \cc \s\uu 2\ ,
\eee
we find that the path integral
over ${\bf b}\ll{[p]}$ and ${\bf a}\ll{([p],\perp)}$ is
${{16\cc \pi\uu 4 }\over{k\sqd \cc |p|\sqd}}$.  The other
contributing factor, from equal and opposite nonzero momenta $[p]$,
is the Fadeev-Popov determinant
${{d\sqd {\bf a}\ll{([p],\parallel)}}\over{d\sqd{\bf c}\ll{[p]}}}$.
The gauge transformation of ${\bf a}\ll{([p],\parallel)}$ is
\bbb
\d \cc {\bf a}\ll{([p],\parallel)} = i
\cc |p|\cc  \cdot \d\cc {\bf c}\ll{[p]}\ , \llsk\llsk
\d \cc {\bf a}\ll{([p],\parallel)}^*
= -i \cc |p|\cc  \cdot \d\cc {\bf c}\ll{[p]}^*\ ,
\eee
so the Jacobian of the gauge transformation is given by
\bbb
{{d\sqd {\bf a}\ll{([p],\parallel)}}\over{d\sqd{\bf c}\ll{[p]}}} = |p|\sqd\ .
\eee
Thus for each equal and opposite pair $[p]$ of nonzero momenta
there is a cancellation of $|p|$-dependence between the dynamical
Gaussian path integral and the Fadeev-Popov determinant,
leaving a factor of $+ {{16\pi\uu 4}\over{k\sqd}}$ for each $[p]$.

Formally, then, the path integral over nonzero modes, modulo the
volume of the group of gauge transformations at nonzero momentum, is
\bbb
Z\ll{\rm nonzero} = \exp{\{F\ll{\rm nonzero}\}}\ ,\llsk\llsk
F\ll{\rm nonzero} = \sum\ll{[p]} F\ll{[p]}\ ,
\nn\\
F\ll{[p]} =  {\rm ln}\lrdd {{16 \pi\uu 4}\over{k\sqd}}
\rrdd = - {\rm ln}\lrdd {{k\sqd}\over{16 \pi\uu 4}} \rrdd\ .
\eee

The set of equal and opposite pairs $[p]$ in the sum above
is indexed by a set of half the momenta.
Since the summand is invariant under $p\to - p$, it is easier
to halve the summand and let the sum run over all momenta:
\bbb
F\ll{\rm nonzero} = \sum\ll{p} F\ll{p}\ ,\llsk\llsk F\ll p \equiv \hh F\ll{[p]}
= + {\rm ln}\lrdd {{4\pi\sqd}\over k}\rrdd
= - {\rm ln}\lrdd{{k}\over{4\pi\sqd}} \rrdd\ .
\eee
Formally, then, the path integral
over the nonzero modes $B$ and $A$,
dividing out by the volume of gauge group at nonzero momenta, is given by
\bbb
Z\ll{\rm nonzero} & = & \exp{\{ - {\cal F}\cdot
{\rm ln}\lrdd{{k}\over{4\pi\sqd}} \rrdd \}}\ ,
\nn\\
{\cal F} & \equiv & \sum\ll {p\neq 0}\cc 1\ .
\eee
The quantity ${\cal F}$ awaits
regularization and renormalization.  For the moment we assert that
${\cal F} = -1 + q \cdot {\rm vol}\ll 2$
for any local renormalization procedure, where
${\rm vol}\ll 2$ is the volume
$4\pi\sqd r\ll 1 r\ll 2 = 2\pi \b r\ll 1$ of the two-torus,
and $q$ is a counterterm adjusting the effective vacuum energy density:
\bbb
q =  \rho\ll \L / {\rm ln}\lrdd {{k}\over{4\pi\sqd}}\rrdd \ .
\eee
The choice of $q$ (or equivalently $\rho\ll\L$) is a local counterterm
and the magnitude of its finite piece is inherently ambiguous
in the absence of some symmetry principle to determine it.  Since the
classical action is scale invariant, we are motivated to choose the
value of $q$ that restores scale-invariance, namely $q=0$.  The magnitude
of the non-extensive piece of ${\cal F}$ cannot be absorbed into a
local counterterm, is unambiguously determined, and
should be the same for any local renormalization procedure to define
${\cal F}$.  Thus we will choose a renormalization proceure of the general form
\bbb
{\cal F} \to {\cal F}\ll{\rm regulated} + (\Delta q) \cdot {\rm vol}\ll 2,
\eee
where $(\Delta q)$ is always chosen so that the
extensive piece of ${\cal F}$ for whatever regulator we choose:
\bbb
\Delta q = - \lim\ll{{\rm vol}\ll 2 \to \infty}
  {{\cal F\ll{\rm regulated}}\over{{\rm vol}\ll 2}}\ .
\eee

Then we exploit the fact that the summand is just a constant $1$,
and thus infrared-finite for small $|p|$, to write ${\cal F}$
as $-1$ plus a counterterm plus a factorized sum,
where each sum depends only on the momentum in a single direction:
\bbb
{\cal F}\ll{\rm regulated}
= -1 + (\Delta q)\cc {\rm vol}\ll 2  + {\cal F}\ll {1,{\rm regulated}}
\cc {\cal F}\ll {2,{\rm regulated}}\ .
\eee
For any local regulator characterized by a scale $\L$, the sums
${\cal F}\ll{1,{\rm regulated}}$ and ${\cal F}\ll{2,{\rm regulated}}$
vanish up to a UV divergent piece proportional to $\L\cc r\ll 1$
(resp. $\L r\ll 2$), as well as terms vanishing more
quickly than $\L\uu{-1}$.  Thus for appropriately
chosen $\Delta q$, the
sum ${\cal F}\ll{\rm regulated}$ goes to $-1$ plus terms that vanish at
least as quickly as a negative power of $\L$ when $\L$ is sent to $\infty$.

The sum ${\cal F}\ll{\rm regulated}$ is
also $-1$ in zeta-function regularization: the term
${\cal F}\ll{1,{\rm regulated}}$ is defined by
\bbb
{\cal F}\ll{1,{\rm regulated}} & \equiv & \sum\ll{p\ll 1}
\cc \m\uu{2s}\cc (g\uu{11} (p\ll 1)\sqd)\uu {-s} =
(\m r\ll 1)\uu{2s}\cc \sum\ll{n\ll 1} n\ll 1 \uu{-2s}
\nn\\ & = &
(\m r\ll 1)\uu{2s}\cc \lrdd 1 + 2 \sum\ll{n\ll \geq 1} n\ll 1\uu{-2s} \rrdd
=  (\m r\ll 1)\uu{2s}\cc \lrdd 1 + 2 \zeta(2 s) \rrdd\ ,
\eee
where $\zeta$ is the Riemann zeta function.  This sum is convergent for
$s > \hh $ but is defined uniquely by analytic continuation for all
values $s\neq 1$.  Removing the regulator corresponds to evaluating
$s = 0$.  Since $\zeta(0) = - \hh$ we have
${\cal F}\ll{1,{\rm regulated}} = 0$, and
likewise ${\cal F}\ll{2,{\rm regulated}} = 0$.  Thus
there is no need for counterterms in
zeta-function regularization: $\Delta q = 0$ and ${\cal F} = -1$.

So we have found that the appropriately renormalized value of ${\cal F}$
is simply $-1$, and thus the renormalized path integral over nonzero modes,
dividing appropriately by the nonzero mode gauge group
measure, is
\bbb
Z\ll{\rm nonzero} = {k\over{4\pi\sqd}}\ .
\eee
The partition function over nonzero modes alone is nonlocal in all
directions, does not have a Hilbert space interpretation and need not be
integer, which is as it should be.  To derive an appropriate Hilbert space
interpretation of the vacuum amplitude,
we need to include the contributions of the zero modes.

\heading{Integral over the zero modes}

We now compute the volume of the
zero modes of $A\ll i$ and $B$, and divide by the volume of the
zero mode of $\chi$.  The measure
for the dynamical zero modes is just given by $\prod\ll{I = 1,2}
d a\ll {0,I} d b\ll 0$ and the measure for
the $\chi$ zero mode is $d c\ll 0$.  For both integrals
the integrand is $1$ and all that remains
is to compute the region of integration.

The zero mode pieces of the
dynamical fields $A\ll i, B$ and the gauge parameter $\chi$ are
\bbb
\left . A\ll i \right | \ll{{\rm zero~mode}} & = &
\sum\ll I \phi\ll {i,(0,I)} \cc a\ll {(0,I)}
\nn\\
\left . B \right | \ll{{\rm zero~mode}} & = & \phi\ll 0 \cc b\ll 0
\nn\\
\left . \chi \right | \ll{{\rm zero~mode}} & = & \phi\ll 0 \cc c\ll 0\ ,
\eee
where $\phi\ll 0$ and $\phi\ll{(0,I)}$
are zero modes of the Laplacian satisfying the orthonormality
conditions.  We take
\bbb
\phi\ll 0 = {1\over{\sqrt{{\rm vol}\ll 2}}} \ , \llsk\llsk
\phi\ll{i,(0,I)} =  \d\ll{iI} \cc \sqrt{{{g\ll{ii}}\over{{\rm vol}\ll 2}}}.
\eee
To determine the fundamental region for the
Wilson lines $a\ll {(0,I)}$, recall that the Wilson lines
are identified under large gauge transformations,
\bbb
\left . A\ll i \right | \ll{{\rm zero~mode}}
\sim A\ll i + {{2\pi}\over{\Delta x\uu i}}\ ,
\eee
where $\Delta x\uu i$ is the
extent of the coordinate $x\uu i$.  In terms of the orthonormalized zero
modes, this translates into
\bbb
a\ll {(0,I)} \sim a\ll{(0,I)} + {{2\pi\cc \d\ll{iI}\cc
\sqrt{{\rm vol}\ll 2} }\over{\sqrt{g\ll {ii}} \Delta x\uu i}} =
{{2\pi\cc \d\ll{iI}\cc \sqrt{{\rm vol}\ll 2}}\over{L\ll i}}\ ,
\eee
where $L\ll i \equiv \sqrt{g\ll{ii}}\Delta x\uu i$
is the physical length of the cycle in the $x\uu i$
direction.

As for the zero modes of $B$ and $\chi$, both are identified under $2\pi$,
and their zero modes have the same normalization,
so their zero modes have the same identifications:
\bbb
b\ll 0 \sim b\ll 0 + 2\pi\cc \sqrt{{\rm vol}\ll 2}\ , \llsk\llsk\llsk
c\ll 0 \sim c\ll 0 + 2\pi\cc \sqrt{{\rm vol}\ll 2}\ .
\eee
So the total partition function for the zero modes is
\bbb
Z\ll{\rm zero} = {{Z\ll{a\ll 0} \cc Z\ll{b\ll 0}}\over{Z\ll{c\ll 0}}}\ .
\eee
The $b\ll 0$ and $c\ll 0$ zero mode integrals are the same,
\bbb
Z\ll {b\ll 0} = \int \ll 0 \uu {{2\pi}
\over{\sqrt{{\rm vol}\ll 2}}} d\cc b\ll 0 = Z\ll{c\ll 0}
\eee
so their ratio is unity,
and we are left with contributions from the $a\ll 0$ zero modes:
\bbb
Z\ll{\rm zero} = Z\ll {a\ll 0} =
   \int \ll 0 \uu {{ {{2\pi\cc  \sqrt{{\rm vol}\ll 2}}\over{L\ll 1}} }} \cc
 \cc\int \ll 0 \uu {{ {{2\pi\cc  \sqrt{{\rm vol}\ll 2}}\over{L\ll 2}}}}
\cc d\sqd a\ll {(0,I)} = {{4\pi\sqd\cc {\rm vol}\ll 2}\over{L\ll 1 L\ll 2}}\ .
\eee
The product of the lengths of
cycles of a rectangular torus is equal to the volume, $L\ll 1 L\ll 2 =
{\rm vol}\ll 2$, so
\bbb
Z\ll{\rm zero} = 4\pi\sqd\ .
\eee

Thus the total partition function over the torus is the
product of the zero mode and renormalized nonzero mode path integrals:
\bbb
Z = Z\ll{\rm zero} \cdot
Z\ll{\rm nonzero} = (4\pi\sqd) \cdot {{k\over{4\pi\sqd}}} = k,
\eee
independent of $r_1$ and $r_2$.
So we conclude that the BF theory
has exactly $k$ quantum states, all of the same energy, which can be
set exactly to zero by a choice of counterterm for the
two-dimensional vacuum energy density.  We see that the BF theory
at level $k \geq 2$ is the minimal Lagrangian realization of a quantum field
theory with $k$ degenerate vacua -- it is minimal in the sense
that it contains only the degenerate vacuua, and nothing
else.

\heading{Hilbert space interpretation of the vacuum amplitude}

The vacuum amplitude is purely
topological: with the appropriate choice of vacuum energy density counterterm,
the partition function
is independent of $r\ll 1$ and $r\ll 2$.  Interpreting the
Euclidean vacuum amplitude on the torus as a thermal partition function
at temperature $\b\uu{-1} = {1\over{2\pi r\ll 2}}$, we see that the
dimension of the Hilbert space is $k$, and all states have
exactly zero energy.  Note, however, that even
if we had chosen a different value for $\Delta q$,
we would have had $k$ degenerate states, with common
energy $E = 2\pi r\ll 1 \rho\ll\L = 2\pi r\ll 1 (\Delta q) / {\rm ln}
\lrdd {{k}\over{4\pi\sqd}}\rrdd$.

\heading{Spectrum and commutation relations
of local operators and line operators}

The operators in question are
\begin{displaymath}
{\cal O}_n(x) \: \equiv \:
: \exp\left( i n B(x) \right) :
\end{displaymath}
which clearly obey ${\cal O}_n \cdot {\cal O}_m = {\cal O}_{n+m}$.
We also have the Wilson line operators
\begin{displaymath}
W_n \: \equiv \: : \exp\left( i n \oint A \right) :\ ,
\end{displaymath}
with the line integral taken over a spatial cycle.
As we shall show shortly,
the Wilson line operators obey simple equal-time commutation relations with
the local operators $\co\ll m(x)$:
\begin{displaymath}
W_n {\cal O}_m \: = \: \xi^{nm} {\cal O}_m W_n
\end{displaymath}
for $\xi = \exp(-2 \pi i/k)$.
As a result, $W$ and ${\cal O}$ commute like clock and shift operators.

Given the dimension of the Hilbert space and the
fact that the operators $W$ and
${\cal O}$ are invertible, it
follows that the operators $W$ and ${\cal O}$ not only commute
but actually act as
clock and shift operators in the standard $k$-dimensional
representation.  The commutation relations
define an algebra whose
smallest nontrivial representation is $k$-dimensional.  Since the
operators $W$ and ${\cal O}$ are invertible,
their representation on the Hilbert space cannot act
as zero identically,
so their representation must be nontrivial, and the standard $k$-dimensional
representation is the only one that is sufficiently small.

\heading{Commutation relations}

Now we shall compute the commutation relations between the
Wilson line operators and the ${\cal O}_n$.
We will work in timelike gauge:  $A_0 = 0$.
In this case, we can take the action to be
\begin{displaymath}
S \: = \: \frac{k}{2 \pi} \int B \partial_0 A_1
\end{displaymath}
hence the conjugate momenta are
\begin{displaymath}
\pi_{A_1} \: = \: \frac{k B}{2 \pi}, \: \: \:
\pi_B \: = \: - \frac{k A_1}{2 \pi}
\end{displaymath}
so we have the equal-time commutators
\begin{displaymath}
\left[ B(x_1), A_1(x_2) \right] \: = \:
-i \frac{2 \pi}{k} \delta(x_1 - x_2).
\end{displaymath}
From this one immediately derives
\begin{displaymath}
\left[ A_1(x), f(B(y)) \right] \: = \:
\frac{2 \pi i}{k} f'(B(y)) \delta(x-y).
\end{displaymath}
Define
\begin{displaymath}
L \: = \: \oint dx A_1(x)
\end{displaymath}
so that
\begin{displaymath}
\left[ L, f(B) \right] \: = \: \frac{2 \pi i}{k} f'(B)
\end{displaymath}
so in particular
\begin{displaymath}
\left[ L, e^{i \alpha B} \right] \: = \: - \frac{ 2 \pi \alpha}{k} e^{i
\alpha B}
\end{displaymath}
for any constant $\alpha$.
It is then straightforward to compute that
\begin{displaymath}
e^{i \beta L} e^{i \alpha B} e^{- i \beta L} \: = \:
\exp\left( - \frac{2 \pi i  \alpha \beta}{k} \right) e^{i \alpha B}
\end{displaymath}
for any constants $\alpha$, $\beta$.
Thus, in particular,
\begin{displaymath}
W_n {\cal O}_m \: = \:
\exp\left( - \frac{2 \pi i m n}{k} \right) {\cal O}_m W_n.
\end{displaymath}

We have already seen that the dimension of the Hilbert space of states
in this theory is $k$.
Now we see that this $k$-dimensional Hilbert space carries a
minimum-dimensional
representation of the
finite-dimensional analog of the Heisenberg algebra, generated by clock
and shift operators at level $k$,
which are generated by natural local operators and line operators
of the theory.

Concretely, the operator ${\cal O}\ll n$ acts on the Hilbert space as
\begin{displaymath}
{\cal O}_n \: \sim \: \left[
\begin{array}{ccccc}
1 & 0 & 0 & \cdots & 0 \\
0 & \xi\uu n & 0 & \cdots & 0 \\
0 & 0 & \xi^{2n} & \cdots & 0 \\
\vdots & \vdots & \vdots & & \vdots \\
0 & 0 & 0 & \cdots & \xi^{(k-1)n}
\end{array} \right].
\end{displaymath}

Note also that the ${\cal O}\ll n$
exhaust the set of linearly independent local operators, as opposed to
line operators, in the theory,
and therefore that the state-operator correspondence holds in this
theory, despite its unfamiliar features: the
dimension of the Hilbert space is $k$, matching the number of
independent ${\cal O}_n$.

Having established the action
of the ${\cal O}\ll n$ on the $k$-dimensional Hilbert space, it
is then possible from
linear combinations of the
${\cal O}\ll n$ to construct $k$ independent projection
operators that are also local operators,
which confirms the decomposition hypothesis that we have
conjectured to hold in general
for two-dimensional conformal theories not satisfying cluster decomposition
\cite{hhpsa}.
For example, the operator $P\ll 0$
projecting onto states invariant under the continuous $B\to B + \e$ symmetry
is given by
\begin{displaymath}
P_0 \: = \: \frac{1}{k} \sum_{i=0}^{k-1} {\cal O}_i.
\end{displaymath}
Of course projection operators always
exist in any finite-dimensional (or even separable) Hilbert space.
The existence of
the projectors has special significance, indicating the failure of
cluster decomposition, only because they are local operators.

\heading{Direct demonstration of non-cluster-decomposition}

We can also deomonstrate the failure of cluster decomposition directly in this
theory, without
considering Wilson lines; we can simply compute the correlation function of
two local ${\cal O}$-operators made from $B$,
and note that the correlation function is not equal to the product of
one-point functions, even when the spacelike separation between the
operators becomes arbitrarily large.
Correlation functions containing only $B$'s are particularly simple,
receiving only
contributions from the zero modes of $B$ -- nonzero modes do not contribute.
This is because $B$ is a field that enters only linearly in the Lagrangian.
In terms of Feynman diagrams,
the propagator is purely anti-diagonal between $B$ and $A$:
\begin{center}
\begin{picture}(100,20)
\ArrowLine(10,10)(90,10)
\Text(10,8)[r]{$B$}  \Text(92,10)[l]{$A$}
\end{picture}
\end{center}
and there are no interaction vertices.
Thus, the expectation value of a set of $B$ nonzero-modes
is equal to its classical value,
{\it i.e.} the value where all nonzero modes of $B$ are set to
zero, because in the
absence of $A$ modes the external $B$ lines have nowhere to terminate.
This argument does not
apply to zero modes of $B$ because these modes do not have a well-defined
propagator and
there is no diagrammatic calculation of their correlation functions.

One implication is that ${\cal O}_n \cdot {\cal O}_m = {\cal O}_{n+m}$.
In particular, this means that
\begin{displaymath}
\langle {\cal O}_n \rangle \: = \:
\langle {\cal O}_1^n \rangle
\end{displaymath}
and from integrating over the circle of $B$ zero modes, we find that
\begin{displaymath}
\langle {\cal O}_1^n \rangle \: = \:
\langle 1 \rangle \delta_{n, 0 \: {\rm mod} \: k}.
\end{displaymath}

The result above shows directly that
cluster decomposition does not hold in the BF theory at level $k\geq 2$.
If cluster decomposition were to hold, it would mean that
\begin{equation}
\lim_{x \rightarrow \infty}
\langle {\cal O}_1(x) {\cal O}_{-1}(0) \rangle
\: = \:
\lim_{x \rightarrow \infty}
{{\langle {\cal O}_1(x) \rangle
\langle {\cal O}_{-1}(0) \rangle}\over{\langle 1 \rangle}}\ .
\label{counterfactual}\end{equation}
On the other hand, from the results above, we know that
\begin{eqnarray*}
\langle {\cal O}_1(x) {\cal O}_{-1}(0) \rangle
& = & \langle 1 \rangle \\
\langle {\cal O}_{\pm 1}(x) \rangle & = & 0\llsk\llsk ({\rm unless~}k=1)
\end{eqnarray*}
so the property~(\ref{counterfactual}) does not hold for $k\geq 2$:
the operators ${\cal O}\ll{\pm 1}$ are correlated with one another
at arbitrary spacelike
separation, as expected from a
summation over multiple degenerate vacua labeled by expectation values
of the ${\cal O}\ll{\pm 1}$.


\begin{thebibliography}{199}


\addcontentsline{toc}{section}{References}

\bibitem{bag-ed1} E. Witten, J. Bagger, ``Quantization of Newton's constant
in certain supergravity theories,'' Phys. Lett. {\bf B115} (1982)
202-206.

\bibitem{zohar1} Z. Komargodski, N. Seiberg, ``Comments on the Fayet-Iliopoulos
term in field theory and supergravity,''
JHEP {\bf 0906} (2009) 007, {\tt arXiv:  0904.1159}.

\bibitem{zohar2} Z. Komargodski, N. Seiberg, ``From linear SUSY to constrained
superfields,'' JHEP {\bf 0909} (2009) 066,
{\tt arXiv:  0907.2441}.

\bibitem{zohar3} Z. Komargodski, N. Seiberg, ``Comments on supercurrent
multiplets, supersymmetric field theories and supergravity,''
JHEP {\bf 1007} (2010) 017,
{\tt arXiv:  1002.2228}.

\bibitem{butter1} D. Butter, ``Conserved currents and Fayet-Iliopoulos terms
in supergravity,'' {\tt arXiv:  1003.0249}.

\bibitem{zohar4} T. Dumitrescu, Z. Komargodski, M. Sudano,
``Global symmetries and D-terms in supersymmetric field theories,''
{\tt arXiv:  1007.5352}.

\bibitem{nati0} N. Seiberg, ``Modifying the sum over topological sectors and
constraints on supergravity,''
{\tt arXiv:  1005.0002}.

\bibitem{git-sugrav} J. Distler, E. Sharpe, ``Quantization of Fayet-Iliopoulos
parameters in supergravity,'' {\tt arXiv:  1008.0419}.

\bibitem{banks-seib} T. Banks, N. Seiberg, ``Symmetries and strings in
field theory and gravity,'' {\tt arXiv:  1011.5120}.


\bibitem{kps} S. Katz, T. Pantev, E. Sharpe, ``D-branes, orbifolds, and
Ext groups,'' Nucl. Phys. {\bf B673} (2003) 263-300,
{\tt arXiv:  hep-th/0212218}.


\bibitem{nr} T. Pantev, E. Sharpe, ``Notes on gauging noneffective
group actions,'' {\tt arXiv:  hep-th/0502027}.

\bibitem{msx} T. Pantev, E. Sharpe, ``String compactifications on
Calabi-Yau stacks,'' Nucl. Phys. {\bf B733} (2006) 233-296,
{\tt arXiv:  hep-th/0502044}.

\bibitem{glsm} T. Pantev, E. Sharpe, ``GLSM's for gerbes (and other
toric stacks),'' Adv. Theor. Math. Phys. {\bf 10} (2006) 77-121,
{\tt arXiv:  hep-th/0502053}.

\bibitem{hhpsa} S. Hellerman, A. Henriques, T. Pantev, E. Sharpe,
M. Ando, ``Cluster decomposition, T-duality, and gerby CFTs,''
Adv. Theor. Math. Phys. {\bf 11} (2007) 751-818,
{\tt arXiv:  hep-th/0606034}.

\bibitem{cdhps} A. Caldararu, J. Distler, S. Hellerman, T. Pantev,
E. Sharpe, ``Non-birational twisted derived equivalences in abelian GLSMs,''
Comm. Math. Phys. {\bf 294} (2010) 605-645,
{\tt arXiv:  0709.3855}.

\bibitem{tonyme} T. Pantev, E. Sharpe, ``Heterotic strings on gerbes,''
to appear.

\bibitem{karp1} R. Karp, ``On the ${\bf C}^n/{\bf Z}_m$ fractional
branes,'' J. Math. Phys. {\bf 50} (2009) 022304,
{\tt arXiv:  hep-th/0602165}.

\bibitem{karp2} C. Herzog, R. Karp, ``On the geometry of quiver gauge
theories:  Stacking exceptional collections,' {\tt arXiv:  hep-th/0605177}.

\bibitem{me-vienna} E. Sharpe, ``Derived categories and stacks in
physics,'' contribution to the proceedings of the ESI research conference
on homological mirror symmetry (Vienna, Austria, June 2006),
{\tt arXiv:  hep-th/0608056}.

\bibitem{me-tex} E. Sharpe, ``Landau-Ginzburg models, gerbes, and
Kuznetsov's homological projective duality,'' to appear in the proceedings
of {\it Topology, ${\bf C}^*$ algebras, string duality} (Texas Christian
University, May 18-22, 2009).

\bibitem{me-qts} E. Sharpe, ``GLSM's, gerbes, and Kuznetsov's homological
projective duality,'' contribution to the proceedings of
{\it Quantum theory and symmetries 6}, {\tt arXiv:  1004.5388}.


\bibitem{cr} W.-M. Chen, Y.-B. Ruan, ``A new cohomology theory for
orbifold,'' Comm. Math. Phys. {\bf 248} (2004) 1-31,
{\tt arXiv:  math/0004129}.

\bibitem{agv} D. Abramovich, T. Graber, A. Vistoli, ``Gromov-Witten theory
of Deligne-Mumford stacks,'' Amer. J. Math. {\bf 130} (2008) 1337-1398,
{\tt arXiv:  math.AG/0603151}.

\bibitem{cclt} T. Coates, A. Corti, Y. Lee, H. Tseng, ``The quantum orbifold
cohomology of weighted projective spaces,'' Acta. Math. {\bf 202} (2009)
139-193,
{\tt arXiv:  math.AG/0608481}.

\bibitem{mann} E. Mann, ``Orbifold quantum cohomology of weighted
projective spaces,'' J. Alg. Geom. {\bf 17} (2008) 137-166,
{\tt arXiv:  math.AG/0610965}.

\bibitem{vistoli} A. Vistoli, ``Intersection theory on algebraic stacks
and on their moduli spaces,'' Inv. Math. {\bf 97} (1989) 613-670.

\bibitem{gomez} T. Gomez, ``Algebraic stacks,'' Proc. Indian Acad. Sci.
Math. Sci. {\bf 111} (2001) 1-31,
{\tt arXiv:  math.AG/9911199}.

\bibitem{lmb} G. Laumon, L. Moret-Bailly, {\it Champs alg\'ebriques},
Springer, 1999.

\bibitem{hv} K. Hori, C. Vafa, ``Mirror symmetry,'' 
{\tt arXiv:  hep-th/0002222}.

\bibitem{mp} D. Morrison, R. Plesser, ``Towards mirror symmetry as duality
for two-dimensional abelian gauge theories,''
Nucl. Phys. Proc. Suppl. {\bf 46} (1996) 177-186,
{\tt arXiv:  hep-th/9508107}.

\bibitem{vw-dt} C. Vafa, E. Witten, ``On orbifolds with discrete torsion,''
J. Geom. Phys. {\bf 15} (1995) 189-214,
{\tt arXiv:  hep-th/9409188}.

\bibitem{kap-02} A. Kapustin, ``Holomorphic reduction of
${\cal N}=2$ gauge theories, Wilson-'t Hooft operators, and S-duality,''
{\tt arXiv:  hep-th/0612119}.


\bibitem{ajt1} E. Andreini, Y. Jiang, H.-H. Tseng, ``On Gromov-Witten theory
of root gerbes,'' {\tt arXiv:  0812.4477}.

\bibitem{ajt2} E. Andreini, Y. Jiang, H.-H. Tseng, ``Gromov-Witten theory
of product stacks,'' {\tt arXiv:  0905.2258}.

\bibitem{ajt3} E. Andreini, Y. Jiang, H.-H. Tseng, ``Gromov-Witten theory
of etale gerbes, i:  root gerbes,'' {\tt arXiv:  0907.2087}.

\bibitem{t1} H.-H. Tseng, ``On degree zero elliptic orbifold
Gromov-Witten invariants,'' {\tt arXiv:  0912.3580}.

\bibitem{gt1} A. Gholampour, H.-H. Tseng, ``On Donaldson-Thomas invariants
of threefold stacks and gerbes,'' {\tt arXiv:  1001.0435}.

\bibitem{xt1} X. Tang, H.-H. Tseng, ``Duality theorems of \'etale gerbes
on orbifolds,'' {\tt arXiv:  1004.1376}.

\bibitem{ed-anton} A. Kapustin, E. Witten, ``Electric-magnetic duality
and the geometric Langlands program,'' {\tt arXiv:  hep-th/0604151}.

\bibitem{edgl2} E. Witten, ``Mirror symmetry, Hitchin's equations, and
Langlands duality,'' {\tt arXiv:  0802.0999}.

\bibitem{ron-tony} R. Donagi, T. Pantev, ``Langlands duality for Hitchin
systems,'' {\tt arXiv:  math.AG/0604617}.

\bibitem{tonypriv} T. Pantev, private communication.

\bibitem{pouliot} P. Pouliot, ``Chiral duals of non-chiral susy gauge
theories,'' Phys. Lett. {\bf B359} (1995) 108-113,
{\tt arXiv:  hep-th/9507018}.

\bibitem{poul-strass} P. Pouliot, M. Strassler, ``A chiral $SU(N)$ gauge
theory and its non-chiral Spin(8) dual,''
Phys. Lett. {\bf B370} (1996) 76-82,
{\tt arXiv:  hep-th/9510228}.

\bibitem{strassler} M. Strassler, ``Duality, phases, spinors and monopoles
in $SO(n)$ and Spin$(n)$ gauge theories,'' JHEP {\bf 9809} (1998) 017,
{\tt arXiv:  hep-th/9709081}.

\bibitem{strassler2} M. Strassler, ``On phases of gauge theories and the
role of non-BPS solitons in field theory,''
{\tt arXiv:  hep-th/9808073}.

\bibitem{triples} J. de Boer, R. Dijkgraaf, K. Hori, A. Keurentjes, J. Morgan,
D. Morrison, S. Sethi, ``Triples, fluxes, and strings,''
Adv. Theor. Math. Phys. {\bf 4} (2002) 995-1186,
{\tt arXiv:  hep-th/0103170}.

\bibitem{gsw1} M. Green, J. Schwarz, E. Witten, {\it Superstring theory,
volume I}, Cambridge University Press, Cambridge, 1987.

\bibitem{ghmr1} D. Gross, J. Harvey, E. Martinec, R. Rohm, ``Heterotic
string theory I:  the free heterotic string,''
Nucl. Phys. {\bf B256} (1985) 253-284.

\bibitem{ginsparg} P. Ginsparg, ``Applied conformal field theory,''
pp. 1-168 in {\it Fields, strings and critical phenomena}
(Les Houches 1988), North-Holland, Amsterdam, 1990,
{\tt arXiv:  hep-th/9108028}.

\bibitem{vafa-qs} C. Vafa, ``Quantum symmetries of string vacua,''
Mod. Phys. Lett. {\bf A4} (1989) 1615-1626.

\bibitem{ed-k} E. Witten, ``D-branes and K theory,''
JHEP 9812 (1998) 019, {\tt arXiv:  hep-th/9810188}.

\bibitem{me-dist} J. Distler, E. Sharpe, ``Heterotic compactifications with
principal bundles for general groups and general levels,''
Adv. Theor. Math. Phys. {\bf 14} (2010) 335-398,
{\tt arXiv:  hep-th/0701244}.

\bibitem{dsw} M. Dine, N. Seiberg, E. Witten ``Fayet-Iliopoulos terms in
string theory,'' Nucl. Phys. {\bf B289} (1987) 589-598.

\bibitem{ads} J. Atick, L. Dixon, A. Sen, ``String calculation of
Fayet-Iliopoulos D-terms in arbitrary supersymmetric compactifications,''
Nucl. Phys. {\bf B292} (1987) 109-149.

\bibitem{bps} M. Bershadsky, T. Pantev, V. Sadov,
``F-theory with quantized fluxes,'' {\tt arXiv:  hep-th/9805056}.

\bibitem{allenpriv} A. Knutson, private communication.

\bibitem{vafa-geom} C. Vafa, ``Geometric origin of Montonen-Olive
duality,'' {\tt arXiv:  hep-th/9707131}.

\bibitem{stringycs} B. Greene, A. Shapere, C. Vafa, S.-T. Yau,
``Stringy cosmic strings and noncompact Calabi-Yau manifolds,''
Nucl. Phys. {\bf B337} (1990) 1-36.

\bibitem{kresch1} A. Kresch, ``On the geometry of Deligne-Mumford stacks,''
pp. 259-271 in {\it Algebraic geometry (Seattle, 2005)}, Proc. Symp. Pure
Math. {\bf 80}, part 1, Amer. Math. Soc., Providence, Rhode Island, 2009,
also available at
{\tt http://www.math.uzh.ch/fileadmin/user/kresch/publikation/geodm.pdf}.

\bibitem{vs} A. Vilenkin, E. P. S. Shellard, {\it Cosmic strings and other
topological defects}, Cambridge University Press, 1994.

\bibitem{sy} M. Shifman, A. Yung, {\it Supersymmetric solitons},
Cambridge University Press, 2009.

\bibitem{git} D. Mumford, J. Fogarty, and F. Kirwan, {\it Geometric
invariant theory}, third edition, Springer-Verlag, 1994.

\bibitem{newstead} P. E. Newstead, {\it Introduction to moduli problems
and orbit spaces}, Tata Institute of Fundamental Research, Springer-Verlag,
Berlin, 1978.

\bibitem{kirwan} F. Kirwan, {\it Cohomology of quotients in symplectic and
algebraic geometry}, Princeton University Press, 1984.

\bibitem{andreithesis} A. C\u ald\u araru,
``Derived categories of twisted sheaves
on Calabi-Yau manifolds,'' Cornell University Ph.D. thesis, 2000.
Available at {\tt http://www.math.upenn.edu/\~{}andreic}

\bibitem{andrei1} A. C\u ald\u araru, ``Derived categories of twisted
sheaves on elliptic threefolds,'' {\tt arXiv:  math/0012083}.

\bibitem{andrei2} A. C\u ald\u araru, ``Non-fine moduli spaces of
sheaves on K3 surfaces,'' {\tt arXiv:  math/0108180}.

\bibitem{cks} A. C\u ald\u araru, S. Katz, E. Sharpe, ``D-branes, B fields,
and Ext groups,'' Adv. Theor. Math. Phys. {\bf 7} (2004) 381-404,
{\tt arXiv:  hep-th/0302099}.

\bibitem{dist-trieste} J. Distler, ``Notes on ${\cal N}=2$ sigma models,''
pp. 234-256 in {\it String theory and quantum gravity 1992} (Trieste, 1992),
{\tt arXiv:  hep-th/9212062}.

\bibitem{mn1} G. Moore, P. Nelson, ``Anomalies in nonlinear sigma models,''
Phys. Rev. Lett. {\bf 53} (1984) 1519-1522.

\bibitem{mn2} G. Moore, P. Nelson, ``The Etiology of sigma model anomalies,''
Comm. Math. Phys. {\bf 100} (1985) 83-132.

\bibitem{mmn} A. Manohar, G. Moore, P. Nelson, ``A comment on sigma
model anomalies,'' Phys. Lett. {\bf B152} (1985) 68-74.

\bibitem{bci} J. Brodie, P. Cho, K. Intriligator,
``Misleading anomaly matchings?,'' Phys. Lett. {\bf B429} (1998)
319-326, {\tt arXiv:  hep-th/9802092}.

\bibitem{is} K. Intriligator, N. Seiberg, ``Mirror symmetry in
three-dimensional gauge theories,'' Phys. Lett. {\bf B387} (1996)
513-519, {\tt arXiv:  hep-th/9607207}.

\bibitem{malmen} A. Malmendier, ``Donaldson invariants of
${\bf C} {\bf P}^1 \times {\bf C}{\bf P}^1$ and mock theta functions,''
{\tt arXiv:  1008.0175}.

\bibitem{pathintegral}
  J.~Polchinski,
  ``Evaluation of the one loop string path integral,''
  Commun.\ Math.\ Phys.\  {\bf 104} (1986) 37-47.


\end{thebibliography}
\end{document}